\documentclass[]{aastex631}

\shorttitle{Observations of instability-driven nanojets}
\shortauthors{Sukarmadji et al.}

\graphicspath{{./}{figures/}}

\begin{document}

\title{Observations of instability-driven nanojets in coronal loops}

\author[0000-0002-9723-3155]{A. Ramada C. Sukarmadji}
\affiliation{Department of Mathematics, Physics and Electrical Engineering, Northumbria University \\ Newcastle upon Tyne, NE1 8ST, UK}

\author[0000-0003-1529-4681]{Patrick Antolin}
\affiliation{Department of Mathematics, Physics and Electrical Engineering, Northumbria University \\ Newcastle upon Tyne, NE1 8ST, UK}

\author[0000-0002-7863-624X]{James A. Mclaughlin}
\affiliation{Department of Mathematics, Physics and Electrical Engineering, Northumbria University \\ Newcastle upon Tyne, NE1 8ST, UK}

\begin{abstract}

The recent discovery of nanojets by \cite{antolin2020} represents magnetic reconnection in a braided field, thus clearly identifying the reconnection-driven nanoflares. Due to their small scale (500~km in widths, 1500~km in lengths) and short timescales ($<$ 15~s), it is unclear how pervasive nanojets are in the solar corona. In this paper, we present IRIS and SDO observations of nanojets found in multiple coronal structures, namely in a coronal loop powered by a blowout jet, and in two other coronal loops with coronal rain. In agreement with previous findings, we observe that nanojets are accompanied by small nanoflare-like intensity bursts in the (E)UV, have velocities of 150-250~km~s$^{-1}$ and occur transversely to the field line of origin, which is sometimes observed to split. However, we find a variety of nanojet directions in the plane transverse to the loop axis. These nanojets are found to have kinetic and thermal energies within the nanoflare range, and often occur in clusters. In the blowout jet case study, the Kelvin-Helmholtz instability (KHI) is directly identified as the reconnection driver. For the other two loops, we find that both, the KHI and the Rayleigh-Taylor instability (RTI) are likely to be the drivers. However, we find that the KHI and RTI are each more likely in one of the other two cases. These observations of nanojets in a variety of structures and environments support nanojets being a general result of reconnection, that are driven here by dynamic instabilities.  

\end{abstract}


\section{Introduction} \label{sec:intro}

The coronal heating problem is an open question in solar physics that has been investigated for many decades, through wave-based and reconnection-based heating mechanisms. A number of wave heating processes have been suggested to be the cause of the heating, mainly through magnetohydrodynamic (MHD) wave dissipation (\citealt{wentzel1979}; \citealt{Klimchuk2006}; \citealt{vandoorsselaere2020}). In particular, transverse MHD waves are good candidates for coronal heating, due to the large amount of energy that they can carry \citep{uchida1974}. Transverse waves in the solar corona have been identified through transverse loop oscillations (\citealt{aschwanden2002}; \citealt{verwichte2004}; \citealt{vandoorsselaere2009}; \citealt{verwichte2009}), and the small-amplitude that are propagating have also been found to be ubiquitous in the solar corona (\citealt{tomczyk2007}; \citealt{anfinogentov2015}). On the other hand, \cite{parker1988} suggested that the heating in the corona was caused by small-scale reconnection events that he termed nanoflares, releasing energy bursts of $10^{24}$ ergs. Small bursts with this energy range have been reported in very different structures, such as in active region moss \citep{testa2013} and particularly at the footpoints of hot loops \citep{testa2014}. The existence of very hot plasma with temperatures of around 10 MK have also been reported by e.g. \cite{reale2011}, \cite{testa2012}, and \cite{ishikawa2017} in non-flaring regions, which is usually attributed to magnetic reconnection. However, no observations of nanoflare-size energy bursts had indicated a direct link towards heating by magnetic reconnections until recently.

The recent discovery of nanojets by \cite{antolin2020} presents major support for reconnection-driven nanoflare heating that occurs in the solar corona. Nanojets are small-scale bursts oriented transversely to the field line of origin, and they are interpreted as the sideways (magnetic tension-driven) motion produced by small-scale reconnections from small-angle field line misalignments, with the localised brightening corresponding to the reconnection region. They were observed in a loop-like structure by the Interface Region Imaging Spectrograph (\textit{IRIS}) \citep{depontieu2014}, where the loss of stability in a hybrid prominence/coronal rain structure nearby was suspected to be the reconnection driver. These nanojets are seen to either form individually or in clusters. Individual occurences mean that a single nanojet forms with no other nanojet formation surrounding it around similar times (on the order of 1 minute), whereas cluster nanojet means that multiple nanojets occur nearby one another (no more than 1500~km apart) and at similar times. This discovery allows us to identify the nanoflares produced by small-angle reconnection in the corona, and thus to distinguish reconnection-driven heating from wave heating. It was also hypothesised that nanojets should be largely independent of the reconnection driver. Additionally, \cite{chen2020} have also reported similar small bursts with the same characteristics that may be linked to nanojets in a solar tornado; however, we still do not have many observations of these nanojets to date. This is partly due to their recent discovery and therefore categorisation, but also because of how short lived (with a timescale of less than 15~s for a single nanojet, as reported by \citealt{antolin2020}) and small-scale they are (average widths of 500 km, lengths of 1500 km). This places a strict spatial and temporal resolution constraint for their detection, and as a result, poses a challenge to estimate how common they are to understand their role in coronal heating; which is the next natural major question that needs answering. 

Because of their ability to produce small-scale structure and deform the magnetic field, dynamic instabilities such as the Kelvin-Helmholtz Instability (KHI) and the Rayleigh-Taylor Instability (RTI) are known to lead to reconnection. Reconnection from the KHI has been simulated by \cite{nykyri2001} with magnetospheric plasma properties, where the reconnection is driven by the vortex motions of the KHI. This is supported by observations of reconnection due to the KHI on the Earth's magnetopause (\citealt{hasegawa2009}), and in the magnetospheric flank (\citealt{nykyri2006}). It was also found that cross-scale energy transport can occur in the regions where there are the KHI vortices (\citealt{moore2016}), further suggesting that the KHI may play an important role in plasma heating.

In the solar atmosphere, transverse MHD waves found in coronal loops have also been shown to generate the Kelvin-Helmholtz Instability (KHI) through TWIKH rolls (for Transverse Wave-Induced KH rolls), with compelling observational evidence reported by \cite{okamoto2015} and confirmed with simulations by \cite{antolin2015}; who observed oscillatory threads in a prominence with signatures of combined resonant absorption and KHI using \textit{IRIS}. Numerical works discussing the likelihood of KHI occuring has been done for different structures such as in spicules (\citealt{ajabshirizadeh2015}) and rotating magnetized jets (\citealt{zaqarashvili2015}), which was then followed up with observations of KHI in various structures including flares (\citealt{zhelyazkov2015}), spicules (\citealt{kuridze2016}; \citealt{antolin2018}), and a Solar Blowout Jet \citep{li2018}. The KHI reported by \cite{li2018} was observed in the outer edges of the jet, indicated by a sawtooth pattern on the regions where it occurs due to a strong velocity shear of 204~km~s$^{-1}$. \cite{yuan2019} also observed multi-layered KHI in the corona, and suggested that plasma heating can be induced by the KHI. Although the presence of magnetic shear can inhibit the growth of the KHI (\citealt{hillier2019}; \citealt{barbulescu2019}), it also provides non-potential magnetic energy available for heating, which may lead to more explosive reconnection.

Another instability we believe to be the driver of the observed structures in the solar corona is the Rayleigh Taylor Instability (RTI). Upflow plumes were observed in quiescent prominences by \cite{berger2008} and \cite{berger2010}, where the RTI is hypothesised as the underlying cause. This was supported by the simulations in \cite{hillier2011} and \cite{hillier2012}, where similar upflow plumes were found to be generated by the RTI. The RTI is also found in simulations of coronal condensations in weak bipolar magnetic fields by \cite{moschou2015} and \cite{xia2017}, which indicates that they may be found in coronal rain.

The results from the observations and simulations have suggested that the KHI and RTI vortices can twist and deform the magnetic field as well as changing the local magnetic pressure, and a myriad of current sheets and vortices are produced through a turbulent cascade. It is also likely that magnetic reconnection is enhanced in the solar corona due to such wave processes (\citealt{howson2021}). However, it is unclear whether this sort of reconnection is only gradual and confined to the smallest turbulent scales, hence, not affecting the overall topology of the loop. The turbulence established by these instabilities also automatically leads to heating, where in the case of TWIKH rolls, has been found to be enough to cover for radiative losses in the Quiet Sun (\citealt{shi2021}). For these reasons, it has been speculated that the TWIKH rolls may contribute to the heating in the solar corona, or perhaps limited to the Quiet Sun (\citealt{soler2019}; \citealt{diaz2021}). With the increasing number of coronal observations of transverse waves and TWIKH rolls, the presence of reconnection processes in these regions also becomes an important question.

In this paper, we present new \textit{IRIS} and \textit{SDO} observations of nanojets. They are found in multiple solar structures, namely in a solar blowout jet within a coronal loop, and in coronal loops with coronal rain, where they occur in different circumstances from the previous reference. The variety of the structures and environment highlights the fact that nanojets can have different drivers, where they may all lead to the same or similar signatures. In section \ref{sec:observations}, the different structures where we have identified nanojets will be explained. Section \ref{sec:analysis} is on the analysis of the nanojets, covering the observed characteristics of the nanojets. Section \ref{sec:temperature} will then discuss about the temperature, number density, and energies of the nanojets. Finally, the results from the previous sections will be discussed in section \ref{sec:discussion} and followed up with a conclusion in section \ref{sec:conclusion}.

\section{Observations} \label{sec:observations}

\begin{figure*}[t!]
  \centering
  \includegraphics[width=0.75\textwidth]{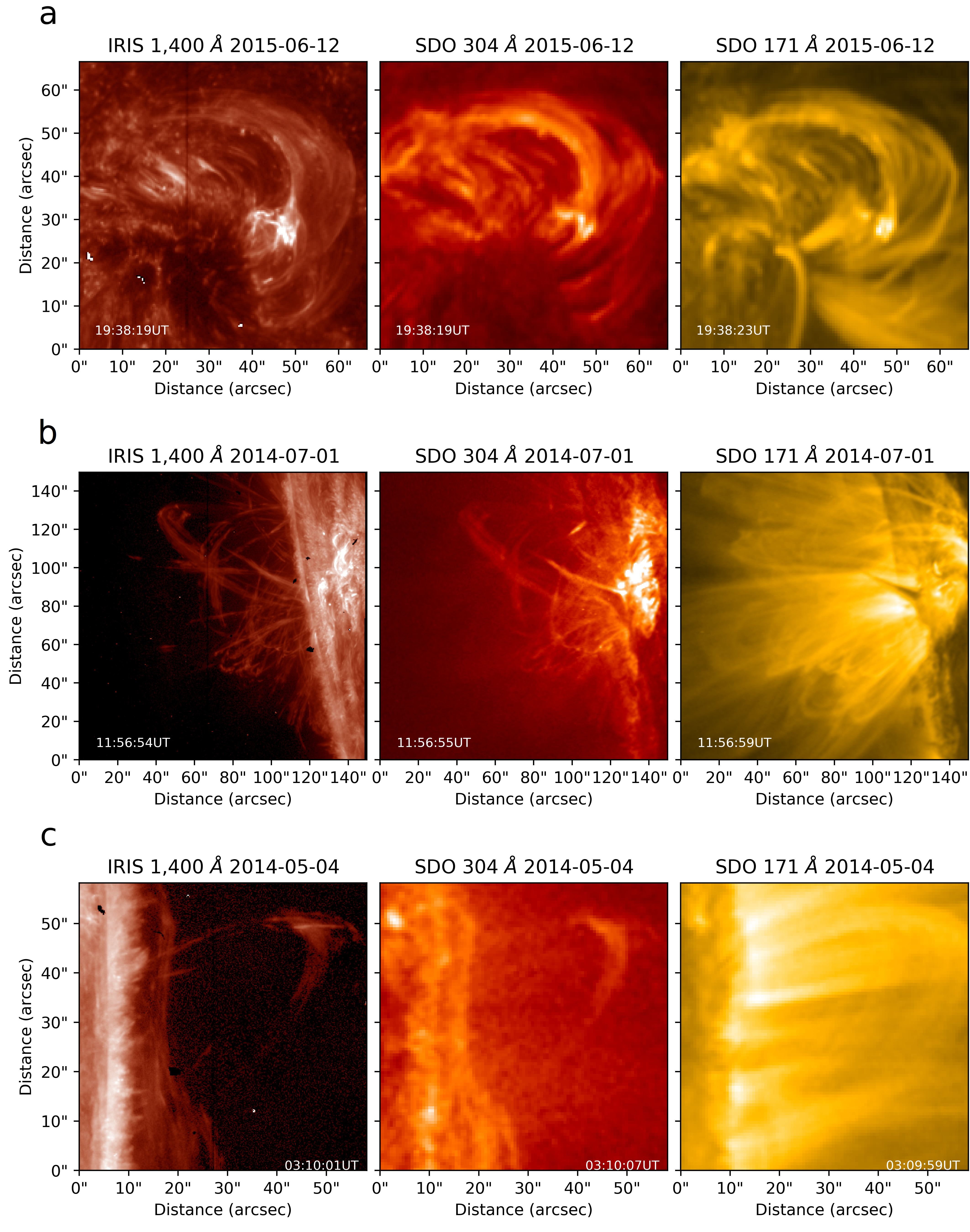}
  \caption{The \textit{IRIS} SJI 1400, AIA 304, and AIA 171 observations of the (a) solar blowout jet, (b) loop structures at east limb, (c) loop structure at west limb.
  \label{fig1}}
\end{figure*}

We will be reporting observations of nanojets in 3 different active regions. All three observations are taken by the \textit{IRIS} and the Atmospheric Imaging Assembly (AIA) from the Solar Dynamics Observatory (\textit{SDO}) (\citealt{lemen2012}).

\subsection{Observation 1: Solar Blowout Jet}
The first observation was taken on the 12th of June 2015, of the active region AR 12365 on disk shown in Figure \ref{fig1}a. The IRIS temporal cadence and platescale for this observation is 3.4 s and 0.33'' respectively, showing a sunspot with a blowout jet captured by the SJI 1400 and AIA. BBSO was observing simultaneously but the structure was not fully visible prior to the event in the H-alpha and AIA EUV channels. 

The jet is seen to form at 19:24 UT, with bright material expanding from the footpoint with both upflows and downflows seen from the cavity. A type of small dome is formed at the footpoint, with the same morphology as null point topologies (\citealt{wyper2018}) where reconnection appears to take place setting off the jet. The jet itself is seen in the SJI 1400 but not in the H-$\alpha$ in BBSO image, meaning that the reconnection is likely taking place near the transition region and that the jet has transition region to coronal temperatures. \cite{li2018} conducted an initial investigation on this same event, focusing on the development of the KHI in the outer edges of the coronal structure following the jet, and reported that the jet is 90 Mm and 19.7 Mm in width. We will be focusing on the outer edges of the coronal structure once it has been disturbed by the jet, where we have found most of the nanojets.

\subsection{Observation 2: Coronal Rain Loops}

For the second dataset we have an observation of an active region on the east limb taken on the 1st of July 2014, with a temporal cadence and platescale of 16.3 s and 0.33'' for IRIS. The active region produced a C class flare with multiple cool flare loops with coronal rain strands visible in the SJI 1330 and 1400, indicating temperature ranges of log(T) = 4.3 - 4.9 (\citealt{depontieu2014}), as well as AIA 304, which has characteristic temperature component of log(T) = 4.7 (\citealt{lemen2012}). The loop-like structure of our focus is formed between 11:20:51 UT and 12:12:54 UT, which is shown in the observations from IRIS and AIA in Figure \ref{fig1}b. As the loop cools and a shower forms, it reaches a maximum height of 52000 km from the surface, with plane-of-sky (POS) plasma flows of around 20-25~km~s$^{-1}$ at the apex and 50-60~km~s$^{-1}$ near the footpoints. 

\subsection{Observation 3: A Loop-Like Structure}

Our last observation was taken on the 3rd of May 2014 at the west limb, of an active region containing multiple loops with quiescent coronal rain with a temporal cadence and platescale of 9.2 s and 0.33'' for the IRIS observation. The main structure for our observations is the loop formed between 02:46:42 - 03:45:36 UT as shown in Figure \ref{fig1}c. During this time period, the loop forms at a height between 26,000 km to 34,000 km with rain and field-aligned flows with POS flow velocities of 17-23~km~s$^{-1}$ at the apex and 50-70~km~s$^{-1}$ near the footpoints. The rain is seen flowing to and from the apex of the loop towards the surface with these velocities. 

This structure is observed in the SJI 2796 and more clearly in the SJI 1400, which suggests a temperature range of log(T) = 4.0 - 4.9. The AIA observations also shows the structure, although they are only seen clearly in the 304 channel and faintly in the 171 channel, which has characteristic temperature components of log(T) = 4.7 and log(T) = 5.6.

\section{Nanojet Analysis} \label{sec:analysis}

\begin{figure*}[t!]
\centering
\includegraphics[width=0.88\textwidth]{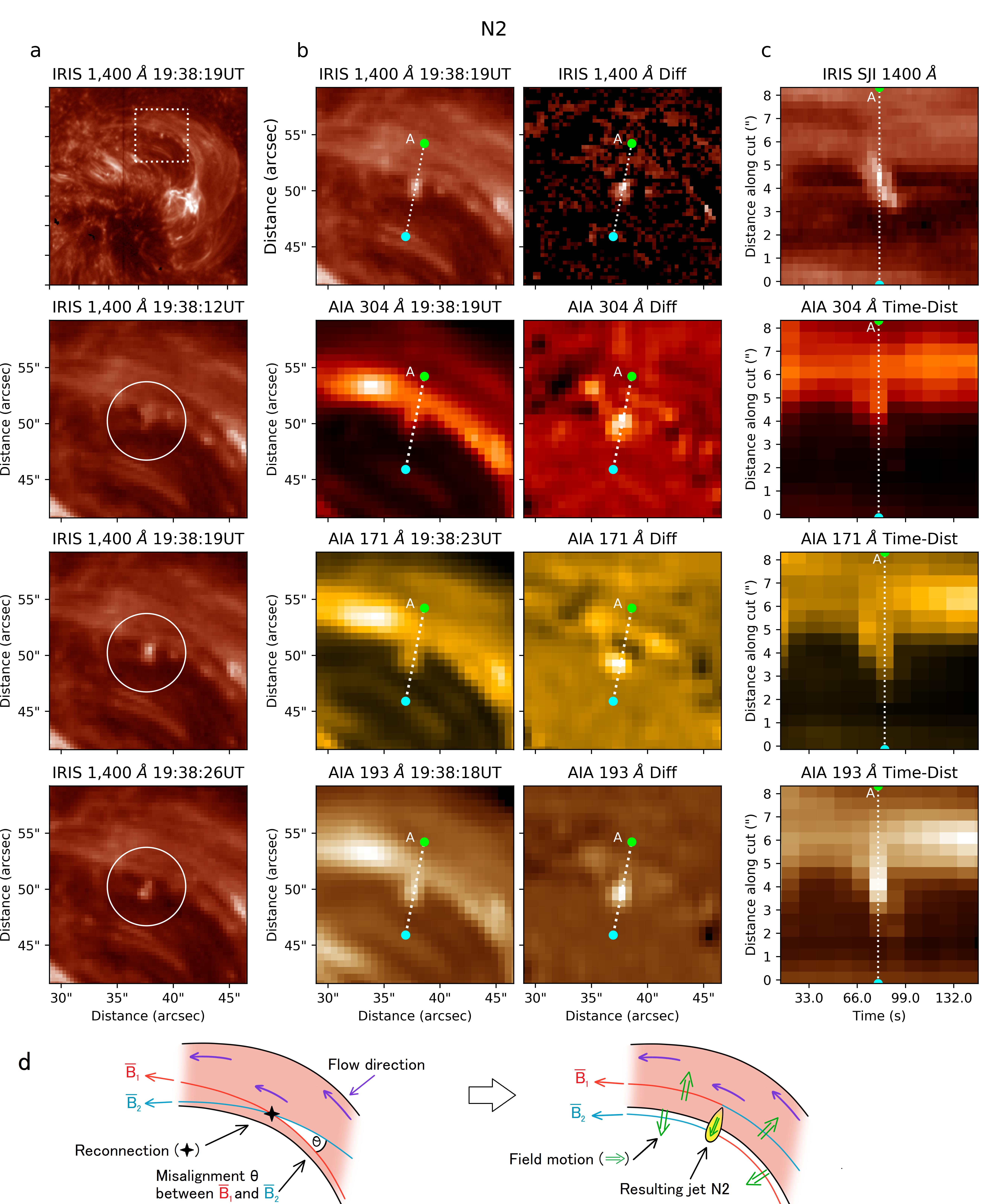}
\caption{A nanojet observed in the solar blowout jet through IRIS SJI 1400, AIA 304, AIA 171, and AIA 193. Plot (a) shows the location of the nanojet (top plot) and its time evolution (the three following plots), (b) is a snapshot sequence showing its time evolution and its corresponding running difference for the snapshot, and (c) is the time-distance diagram of the slice along the nanojet in column (b). The running difference image is constructed by subtracting the current snapshot with the previous snapshot. (d) is a schematic showing how the nanojet is formed and the field configuration during the process. An animation for column (a) is available online, showing the time evolution of the region where the nanojet is found and its location in the blowout jet. 
\label{fig2}}
\end{figure*}

\begin{figure*}[ht]
\centering
\includegraphics[width=0.88\textwidth]{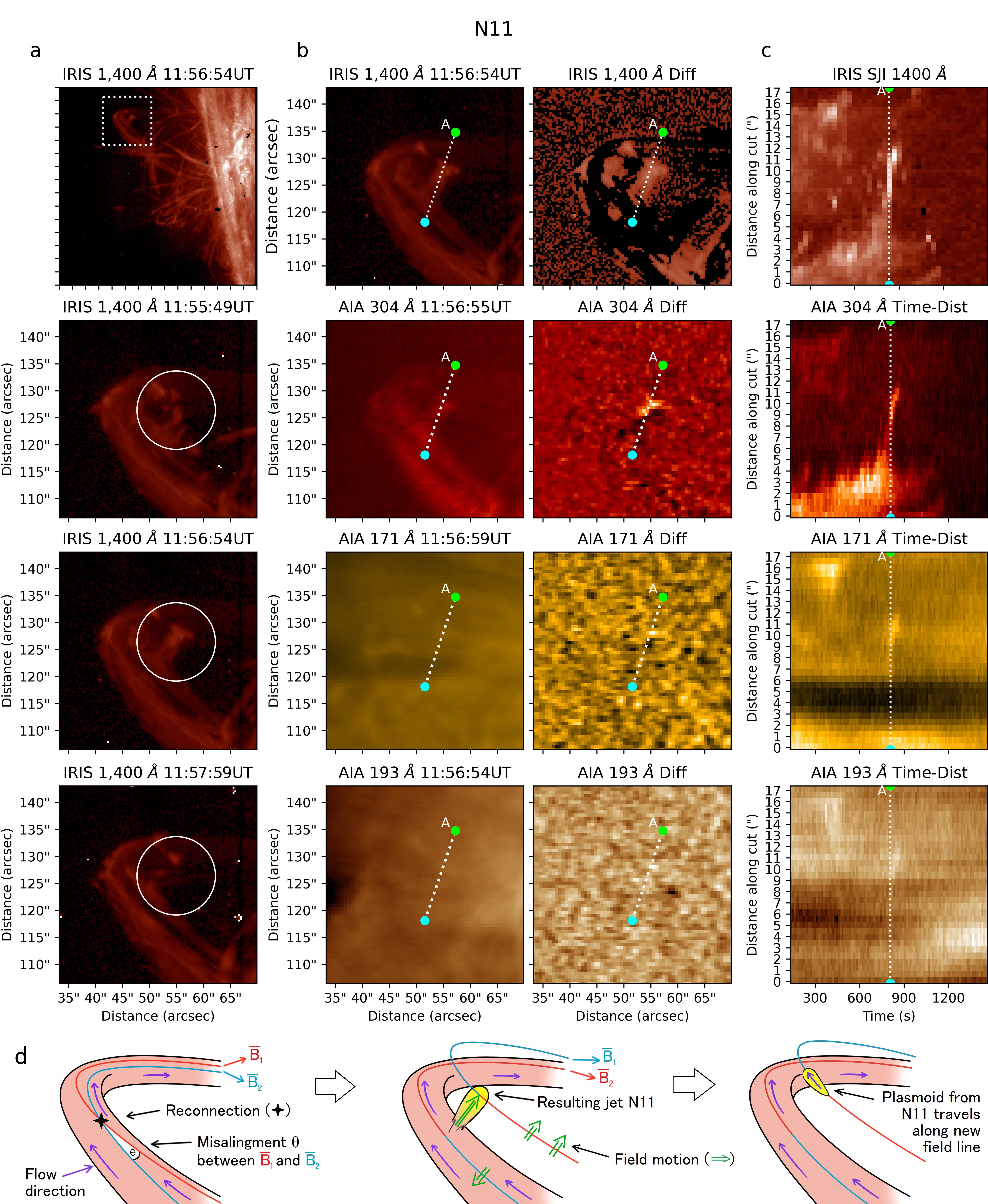}
\caption{A nanojet observed in coronal loops on the E-limb through IRIS SJI 1400, AIA 304, AIA 171, and AIA 193. Column (a) shows the location of the nanojet (top plot) and its time evolution (the three following plots), (b) is a snapshot sequence showing its time evolution and its corresponding running difference for the snapshot, and (c) is the time-distance diagram of the slice along the nanojet in column (b). The running difference image is constructed by subtracting the current snapshot with the previous snapshot. (d) is a schematic showing how the nanojet is formed and the field configuration during the process. An animation for column (a) is available online, showing the time evolution of the region where the nanojet is found and its location in the loop-like structure. 
\label{fig3}}
\end{figure*}

\begin{figure*}[ht]
\centering
\includegraphics[width=0.88\textwidth]{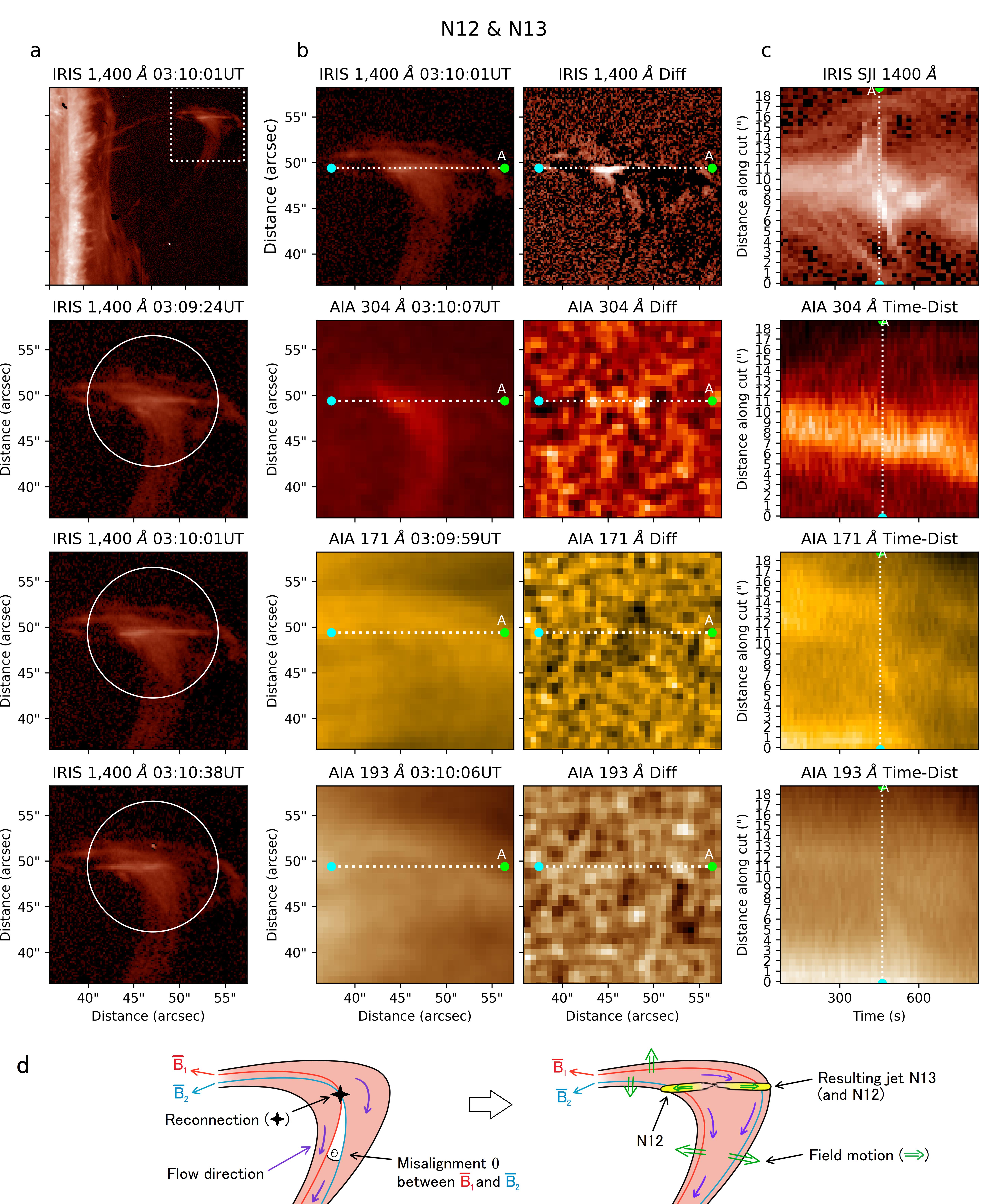}
\caption{A nanojet observed in the W-limb loop structure through SJI 1400, AIA 304, AIA 171, and AIA 193. Column (a) shows the location of the nanojet (top plot) and its time evolution (the three following plots), (b) is a snapshot sequence showing its time evolution and its corresponding running difference for the snapshot, and (c) is the time-distance diagram of the slice along the nanojet in column (b). The running difference image is constructed by subtracting the current snapshot with the previous snapshot. (d) is a schematic showing how the nanojet is formed and the field configuration during the process. An animation for column (a) is available online, showing the time evolution of the region where the nanojet is found and its location in the loop-like structure. 
\label{fig4}}
\end{figure*}

\begin{figure*}[ht]
\centering
\includegraphics[width=0.95\textwidth]{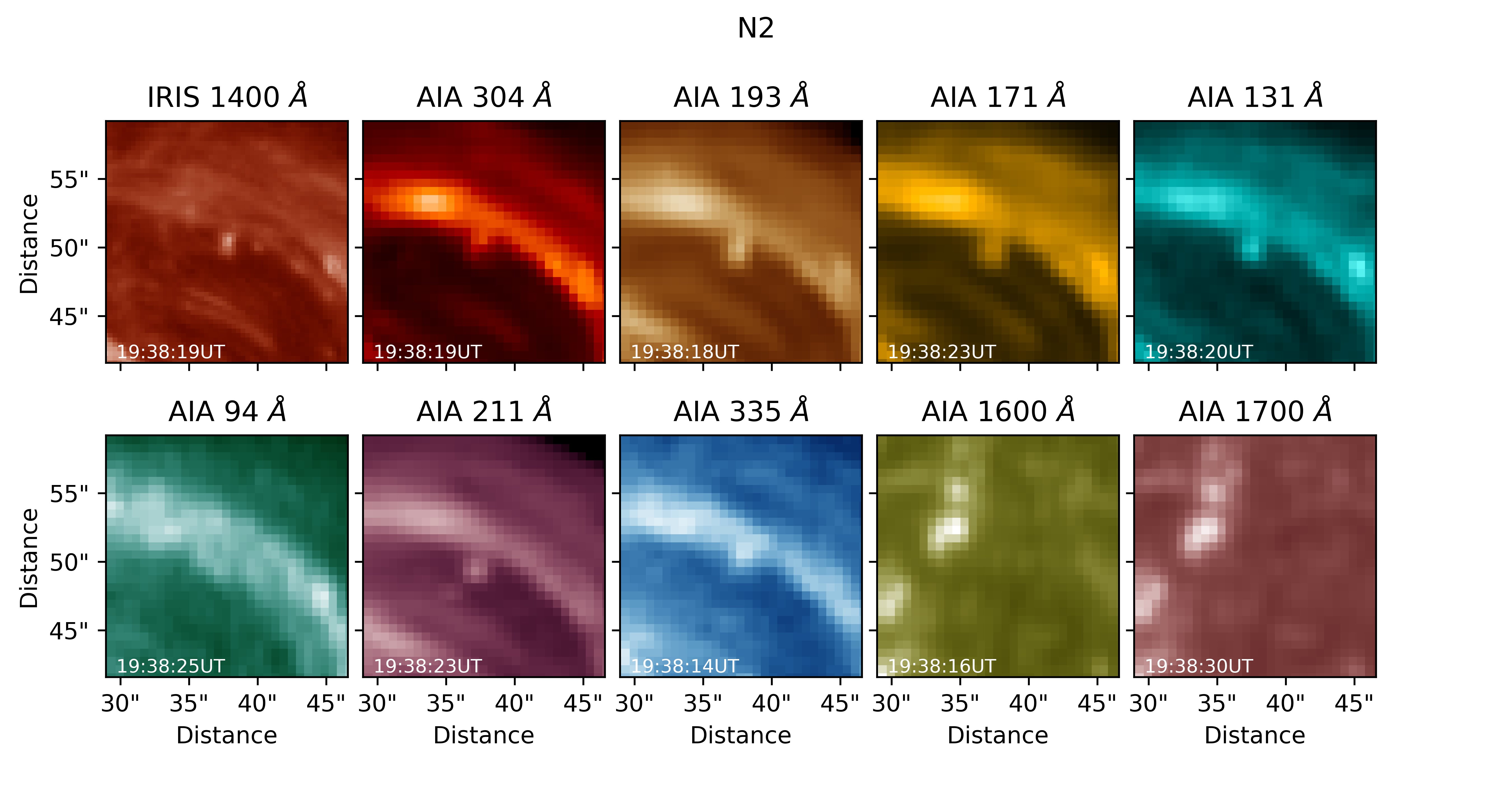}
\caption{A snapshot of a nanojet linked to the blowout jet (N2, same from Figure \ref{fig2}) in different channels, taken at similar times.
\label{fig5}}
\end{figure*}

\begin{figure*}[ht]
\centering
\includegraphics[width=1\textwidth]{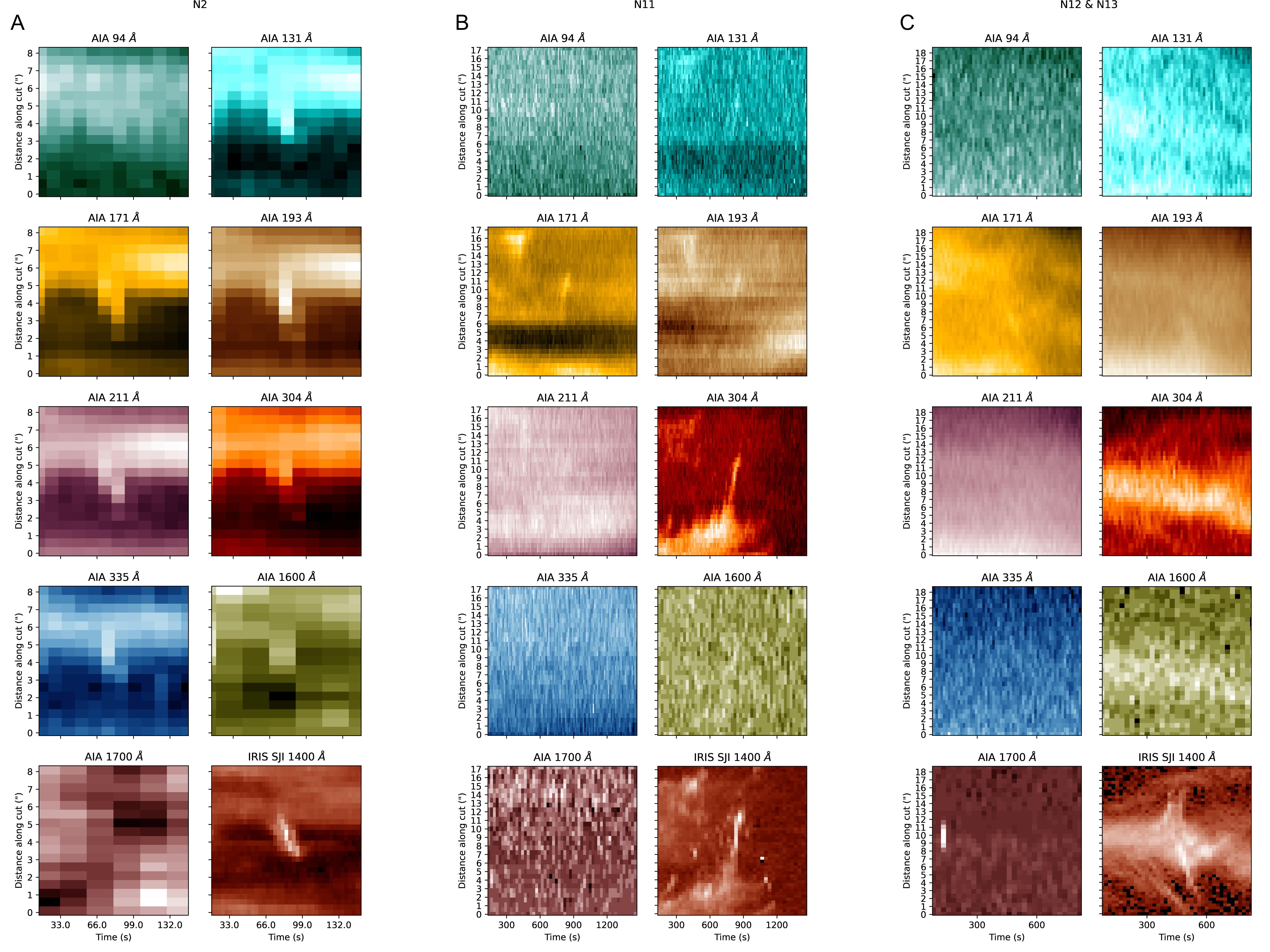}
\caption{Time distance plots at different channels of the image slices taken from Figure 2 of nanojet N2 (A, left), Figure 3 of nanojet N11 (B, middle), and Figure 4 of nanojet N12 and N13 (C, right). 
\label{fig6}}
\end{figure*}

Figures \ref{fig2}-\ref{fig4} shows 1-2 samples of the nanojets found in each of the observations. For the blowout jet observation, the high temporal cadence has allowed us to identify 15 nanojets and observe their evolution as they form and disappear. The nanojets, which are characterised by small bursts of plasma, form after the blowout jet has pervaded the coronal loop, 12 minutes after it starts, when flows along the structure have settled to 100 - 200~km~s$^{-1}$. Prior to a nanojet occurrence, we typically observe a brightening in the location where it forms, and this is then followed by a small ejection (the nanojet) that is closely perpendicular to the field line. An example of a nanojet is shown in Figure \ref{fig2}, where a nanojet is formed at the edges of the coronal loop subject to the blowout jet which is ejected perpendicularly to the field line of origin. We observe that the perpendicular direction of the ejection is a characteristic behaviour of the nanojet and is seen in many of the instances. Moreover, the nanojets are seen in the \textit{IRIS} SJI 1400, in most of the AIA wavelength channels, only very faintly in 94, but no signatures in 1700. An example for this is shown in the image plots in Figure \ref{fig5}, as well as the time distance plots in Figure \ref{fig6}.

\begin{figure*}[ht]
\centering
\includegraphics[width=0.8\textwidth]{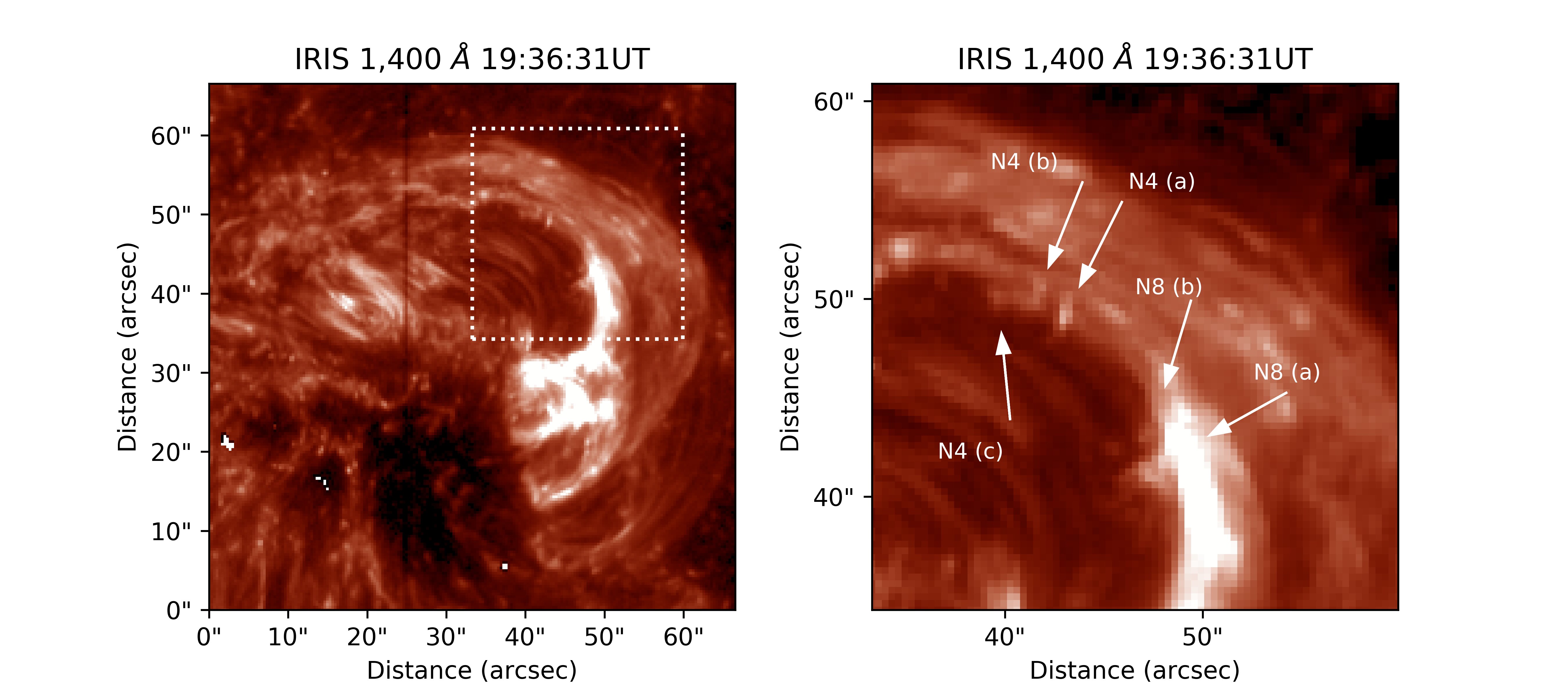}
\caption{An example of nanojets forming in 2 clusters N4 and N8. Cluster N4 contains 3 nanojets (N4(a), N4(b), and N4(c)), and cluster N8 contains 2 nanojets (N8(a), N8(b)), where the nanojets in each cluster are seen forming neighbouring each other. The video showing the time evolution of this plot is also available online, which spans before the nanojet forms and until it has disappeared.
\label{fig7}}
\end{figure*}

Most of the nanojets are seen to be formed at the edges of the blowout jet structure. Some of the nanojets are seen as single events individually, and some of them are found in clusters. For the individual nanojet instances, this means that they occur alone without any other nanojets forming within its surroundings. Whereas the nanojets occurring in clusters form neighbouring each other within 1500~km and 10~s apart from one another. An example of cluster nanojets is shown in Figure \ref{fig7}. Here, we observe clusters N4 and N8 containing 3 and 2 nanojets respectively (N4(a), N4(b), N4(c), and N8(a) and N8(b)). These are the most visible nanojets within the two clusters, but it must be noted that there are other nanojets as well within this cluster. We also observe instances of successive nanojets forming in the same location, within less than 15~s after a previous nanojet, in the same location within less than 15~s after the first nanojet. 

In the second dataset (loop structure on the E-limb) we only observed one occurrence of a nanojet which is formed at the apex of the loop, shown in Figure \ref{fig3}. The nanojet here is most visible in the SJI 1400 and AIA 304, and only faintly visible in the time distance diagrams of AIA 171, 193, and 131 (also shown in Figure \ref{fig6}). We were also able to clearly observe an ejected plasmoid-like structure from the nanojet stopping at a neighboring field line position after its perperndicular travel (as if captured by a different field line), followed again by a change in its motion to follow the field the line (motion parallel to the initial coronal rain flow) and changing its motion to follow the field line. For this observation, we also observe field line splitting in the loop-like structure shown in Figure \ref{fig10}. The loop initially appears to be uniform in intensity, but splits into two strands just before the nanojet is ejected.

As for the third observation, we observed 4 nanojets with the first 2 nanojets seen to be forming around the apex of the loop shown in Figure \ref{fig4}. This first set of nanojets forms 23 minutes after the loop appears, and the second set of nanojets, which formed in clusters in the apex of the loop, followed after and are ejected towards the solar surface. The first two nanojets are ejected simultaneously in opposite directions. One is ejected upwards closely perpendicular from the field line in the plane-of-the-sky, while the other is ejected in the opposite direction back towards the initial footpoint, shown in Figure \ref{fig4}b. In the observation, the nanojet is only most visible in SJI 1400 and faintly in AIA 304. The time distance plot of Figure \ref{fig4}c and the right side of Figure \ref{fig6} shows a slice through the two nanojets, and we observe a herring-bone pattern in the plot which is characteristic of bi-directional jets showing two different directions of plasma ejection. The nanojet ejected downwards is aligned in the plane-of-the-sky with the loop at the footpoint. However, it must be noted that the LOS of the observation makes an oblique angle with the loop plane, so we suspect that this downward nanojet is also closely perpendicular to the local field.

\begin{figure*}[t!]
\centering
\includegraphics[width=1\textwidth]{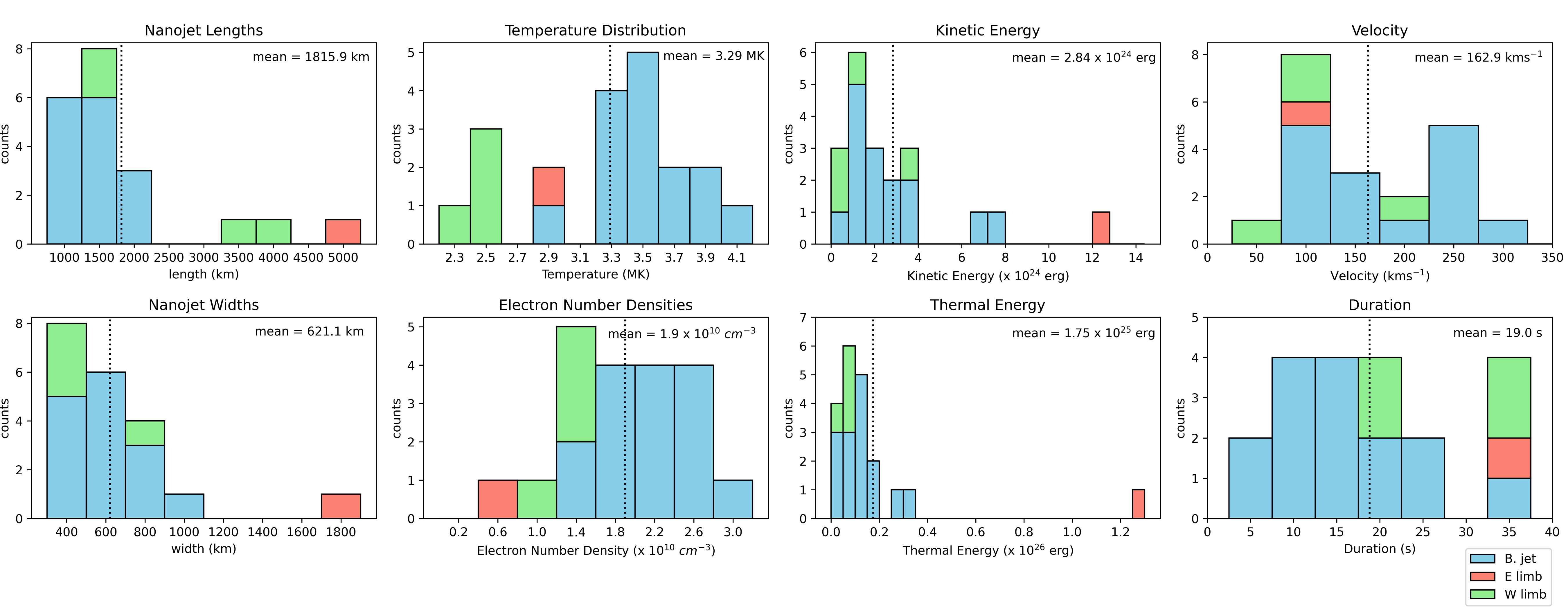}
\caption{Histograms for the lengths, widths, electron number densities, temperatures, kinetic energies, thermal energies, velocities, and duration of the nanojets from all three observations.
\label{fig8}}
\end{figure*}

We measured the sizes of the nanojets by selecting the pixels that shows an intensity contrast from its surrounding, and measured them 6 times to provide a measure of the error involved. The length and width is measured at the snapshot where the nanojet is at its largest, where the length is defined as the longest line parallel to the direction of travel that we can draw on the nanojet, and the width is the longest line perpendicular to the length that we can draw on the nanojet. For the smaller or fainter nanojets, we select the pixels only if it is also visible in the running difference image of the frame where we see the nanojet and the previous frame. From all the observations combined, the average length and widths of the nanojets are 1815 $\pm$ 133 km and 621 $\pm$ 107 km respectively. A histogram of the lengths and widths are shown in Figure \ref{fig8}, and we observe that the length of the nanojets are mostly between 750-2250~km, with three nanojets having lengths over 3000~km. Most of the widths are also found in the range between 300-1100~km, with one instance found having a width of around 1800~km. The largest nanojet is the nanojet found on the E-limb loops (dataset 2) with a length of ~5000~km and 1800~km, which is an outlier in comparison to the other nanojets. Whereas the nanojets on the W-limb loop have average to larger sizes (1500-3800~km in length and 420-720~km in width), and the nanojets in the blowout jet observations are typically shorter but with average widths (800-2200~km in length and 360-990~km in width). Full details of all the nanojet's properties are found in Table \ref{table1}. It is also important to note that from all of the nanojet observations, we only see them in areas near the apex, where curvature is clearly seen.

\begin{figure*}[t]
\centering
\includegraphics[width=0.9\textwidth]{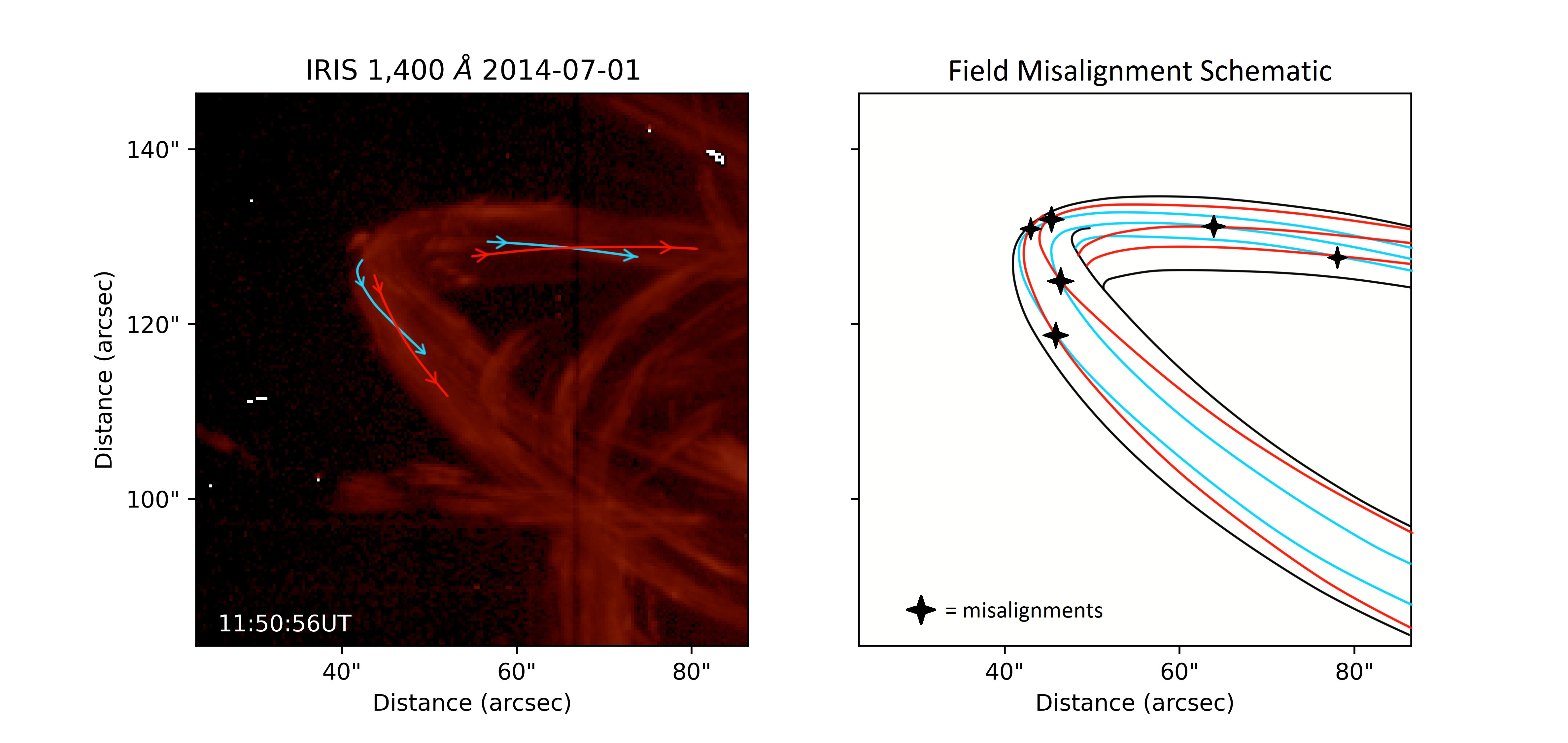}
\caption{A snapshot of the loop with lines indicating the flow misalignments seen in data 2 (left), and a schematic showing the possible misalignments in the structure.
\label{fig9}}
\end{figure*}

\begin{figure*}[t!]
\centering
\includegraphics[width=0.9\textwidth]{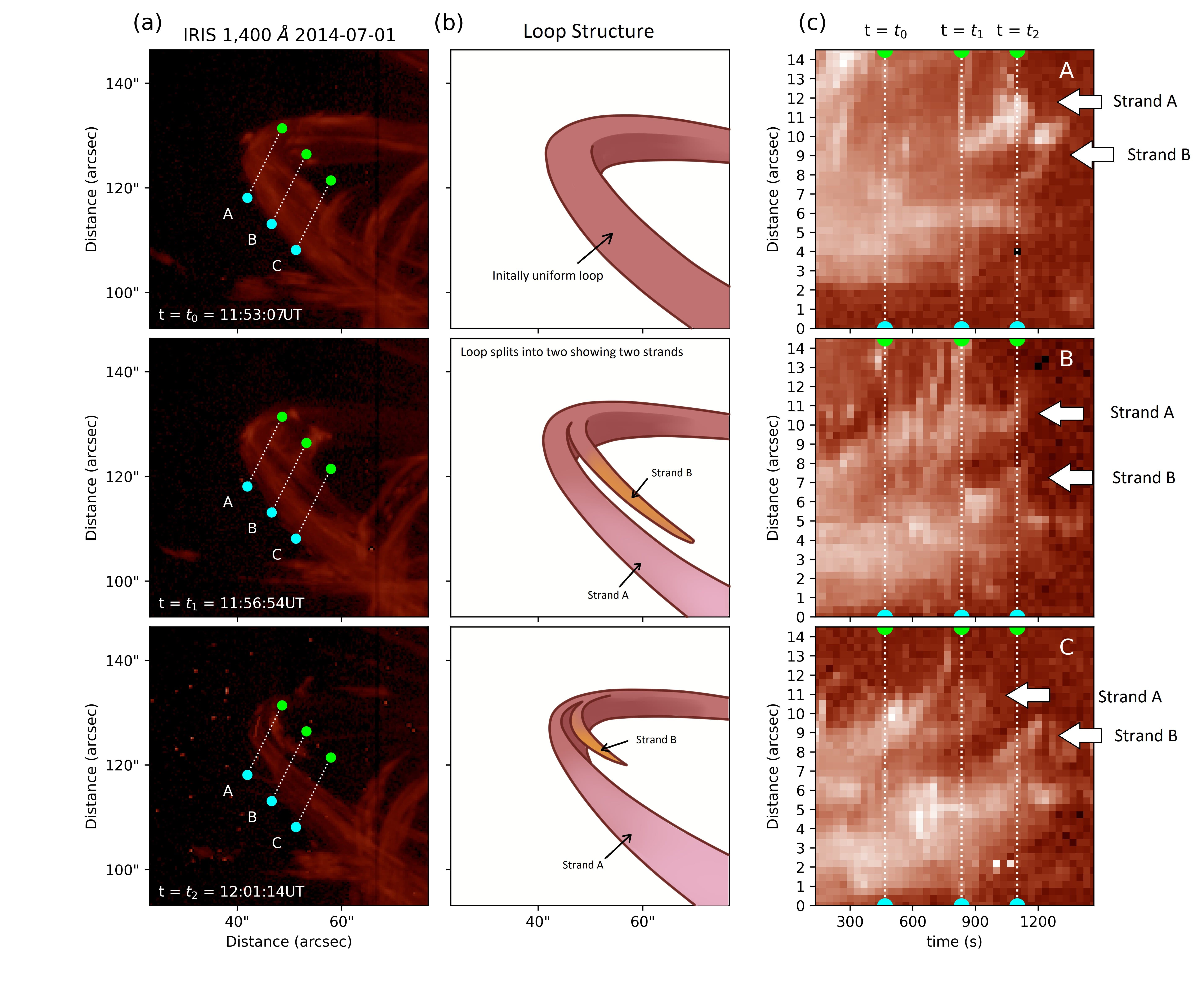}
\caption{An example of field splitting observed in the structure from data 2. (a) shows snapshots from IRIS of the structure at 3 different times (top to bottom, at $t=t_1, t_2, t_3$), with slices A, B, and C that cuts through the structure. (b) is a schematic showing how the structure appears in IRIS during its respective snapshots on the left, which shows how the initially uniform loop eventually splits into strand A and B. The slices A, B, and C from the snapshots in column (a) are plotted as time-distance diagrams in column (c) (top to bottom), with the vertical dotted lines indicating the time of the snapshots from (a). The loop initially appears uniform as indicated by a uniformly bright area at the start of diagrams, but then starts to split around $t=t_1$ into strands A and B, as indicated by the two bright lines that comes from the initially uniform structure.
\label{fig10}}
\end{figure*}

An additional observation that is also worth noting is that we observe appearingly misaligned flows in the loop structures. An example of this is shown in Figure \ref{fig9}, where in the E-limb loop, there are two flows (red and blue lines) that have projected misalingments angles between 5$^{\circ}$-10$^{\circ}$. The nanojet in this observation is also followed by field splitting which is indicated by an initially uniform distribution of coronal rain in the loop followed by rain only flowing in separate, parallel field lines, shown in Figure \ref{fig10}. In the first row of column (a) of Figure \ref{fig10}, the rain initially cover the whole loop apex at $t = t_0$, but at $t = t_{1}$ when the nanojet has formed, the loop appears to have split into two strands (strand A and B). The rain flow only continues in these two strands until the loop completely vanishes, and therefore two lines of intensity for each strands are seen in the three time-distance slices of Figure \ref{fig10}c, where there is a dark gap between the two bright lines indicating the strands.

\begin{deluxetable*}{cccccccccc}
\tablenum{1}
\tablecaption{Nanojet properties from the datasets (Data) showing its length (l), width (w), electron number density ($n_{e}$), temperature (T), velocity (v), kinetic energy (KE), thermal energy (TE), and duration (t). The uncertainties associated with the mean values of each quantity is the error mean. Some of the nanojets from the W-limb have large and unphysical errors from the DEM, but it must be noted that the nanojets are still visible in the coronal channels which suggests that they are within coronal temperatures.\label{table1}}
\tablewidth{0pt}
\tablehead{
\colhead{Data} & \colhead{ID} & \colhead{l} & \colhead{w} & \colhead{$n_{e}$} & \colhead{T}  & \colhead{v} & \colhead{KE} & \colhead{TE} & \colhead{t} \\
\colhead {} & \colhead{} & \colhead{($x10^{3}$~km)} & \colhead{($x10^{2}$~km)} & \colhead{($x10^{10}$~cm$^{-3}$)} & \colhead{(MK)} & \colhead{(~km~s$^{-1}$)} & \colhead{($x10^{24}$~erg)} & \colhead{($x10^{25}$~erg)} & \colhead{(s)}}
\startdata
B. Jet & 1 & 2.2 $\pm$ 0.2 & 6.4 $\pm$ 1.3 & 1.7 $\pm$ 0.01 & 3.4 $\pm$ 0.2 & 101 $\pm$ 22 & 1.5 $\pm$ 0.9 & 2.0 $\pm$ 1.0 & 13.5 $\pm$ 1.7 \\ 
 & 2 & 1.5 $\pm$ 0.1 & 9.9 $\pm$ 1.2 & 1.4 $\pm$ 0.02 & 3.4 $\pm$ 0.5 & 102 $\pm$ 14 & 2.1 $\pm$ 0.7 & 2.7 $\pm$ 1.0 & 20.3 $\pm$ 1.7 \\ 
 & 3 & 1.8 $\pm$ 0.1 & 3.6 $\pm$ 1.2 & 2.7 $\pm$ 0.01 & 3.5 $\pm$ 0.1 & 170 $\pm$ 34 & 1.7 $\pm$ 1.4 & 0.9 $\pm$ 0.6 & 6.8 $\pm$ 1.7 \\ 
 & 4a & 1.4 $\pm$ 0.2 & 5.1 $\pm$ 1.7 & 2.0 $\pm$ 0.01 & 3.3 $\pm$ 0.2 & 232 $\pm$ 78 & 3.6 $\pm$ 3.4 & 0.9 $\pm$ 0.7 & 23.7 $\pm$ 1.7 \\ 
 & 4b & 1.5 $\pm$ 0.1 & 6.4 $\pm$ 1.0 & 1.9 $\pm$ 0.01 & 3.8 $\pm$ 0.1 & 268 $\pm$ 31 & 7.7 $\pm$ 3.3 & 1.7 $\pm$ 0.7 & 13.5 $\pm$ 1.7 \\ 
 & 4c & 1.0 $\pm$ 0.1 & 7.1 $\pm$ 1.0 & 1.6 $\pm$ 0.01 & 4.1 $\pm$ 0.2 & 116 $\pm$ 18 & 1.0 $\pm$ 0.4 & 1.2 $\pm$ 0.5 & 10.1 $\pm$ 1.7 \\ 
 & 5a & 1.6 $\pm$ 0.1 & 3.6 $\pm$ 1.2 & 2.8 $\pm$ 0.01 & 3.6 $\pm$ 0.1 & 155 $\pm$ 36 & 1.3 $\pm$ 1.0 & 0.8 $\pm$ 0.6 & 13.5 $\pm$ 1.7 \\ 
 & 5b & 1.0 $\pm$ 0.2 & 6.4 $\pm$ 0.7 & 2.1 $\pm$ 0.01 & 3.7 $\pm$ 0.1 & 286 $\pm$ 50 & 6.8 $\pm$ 3.0 & 1.2 $\pm$ 0.5 & 10.1 $\pm$ 1.7 \\ 
 & 5c & 0.9 $\pm$ 0.2 & 7.5 $\pm$ 0.7 & 1.8 $\pm$ 0.005 & 3.8 $\pm$ 0.1 & 196 $\pm$ 18 & 3.1 $\pm$ 1.3 & 1.2 $\pm$ 0.5 & 10.1 $\pm$ 1.7 \\ 
 & 6 & 0.8 $\pm$ 0.1 & 3.6 $\pm$ 1.2 & 2.6 $\pm$ 0.12 & 3.8 $\pm$ 1.4 & 241 $\pm$ 73 & 1.5 $\pm$ 1.4 & 0.4 $\pm$ 0.4 & 10.1 $\pm$ 1.7 \\ 
 & 7 & 1.6 $\pm$ 0.2 & 5.1 $\pm$ 1.7 & 2.0 $\pm$ 0.01 & 3.3 $\pm$ 0.2 & 79 $\pm$ 8 & 0.5 $\pm$ 0.4 & 1.1 $\pm$ 0.8 & 23.7 $\pm$ 1.7 \\ 
 & 8a & 2.2 $\pm$ 0.2 & 8.6 $\pm$ 1.7 & 1.7 $\pm$ 0.01 & 3.0 $\pm$ 0.2 & 117 $\pm$ 44 & 3.5 $\pm$ 2.5 & 3.1 $\pm$ 1.5 & 33.8 $\pm$ 1.7 \\ 
 & 8b & 1.4 $\pm$ 0.2 & 5.1 $\pm$ 1.7 & 2.3 $\pm$ 0.01 & 3.5 $\pm$ 0.1 & 126 $\pm$ 37 & 1.2 $\pm$ 1.1 & 1.1 $\pm$ 0.9 & 20.3 $\pm$ 1.7 \\ 
 & 9 & 0.8 $\pm$ 0.1 & 3.6 $\pm$ 1.2 & 2.4 $\pm$ 0.01 & 3.6 $\pm$ 0.2 & 268 $\pm$ 89 & 1.8 $\pm$ 1.7 & 0.4 $\pm$ 0.3 & 6.8 $\pm$ 1.7 \\ 
 & 10 & 1.1 $\pm$ 0.1 & 3.6 $\pm$ 1.2 & 2.7 $\pm$ 0.01 & 3.4 $\pm$ 0.1 & 259 $\pm$ 59 & 2.4 $\pm$ 2.0 & 0.5 $\pm$ 0.4 & 16.9 $\pm$ 1.7 \\ 
E-limb & 11 & 5.0 $\pm$ 0.1 & 18.5 $\pm$ 1.0 & 0.7 $\pm$ 0.02 & 2.8 $\pm$ 0.8 & 105 $\pm$ 34 & 12.3 $\pm$ 5.9 & 12.9 $\pm$ 4.2 & 32.5 $\pm$ 16.3 \\ 
W-limb & 12 & 3.6 $\pm$ 0.1 & 4.2 $\pm$ 0.6 & 1.4 $\pm$ 0.5 & 2.5 $\pm$ 3.6 & 54 $\pm$ 3 & 0.2 $\pm$ 0.1 & 0.8 $\pm$ 1.3 & 36.8 $\pm$ 9.2 \\ 
 & 13 & 3.8 $\pm$ 0.1 & 4.5 $\pm$ 0.4 & 1.4 $\pm$ 0.49 & 2.5 $\pm$ 7.2 & 188 $\pm$ 11 & 3.5 $\pm$ 1.4 & 1.0 $\pm$ 2.9 & 36.8 $\pm$ 9.2 \\ 
 & 14a & 1.6 $\pm$ 0.1 & 7.2 $\pm$ 0.4 & 1.0 $\pm$ 0.24 & 2.3 $\pm$ 4.1 & 102 $\pm$ 3 & 0.8 $\pm$ 0.2 & 0.7 $\pm$ 1.4 & 18.4 $\pm$ 9.2 \\ 
 & 14b & 1.5 $\pm$ 0.1 & 3.9 $\pm$ 0.3 & 1.4 $\pm$ 0.16 & 2.6 $\pm$ 1.8 & 95 $\pm$ 3 & 0.3 $\pm$ 0.1 & 0.3 $\pm$ 0.2 & 18.4 $\pm$ 9.2 \\ 
 \cline{1-10}
 Mean &  & 1.8 $\pm$ 0.1 & 6.2 $\pm$ 1.1 & 1.9 $\pm$ 0.1 & 3.3 $\pm$ 1.1 & 163 $\pm$ 33 & 2.8 $\pm$ 1.6 & 1.7 $\pm$ 1.0 & 18.8 $\pm$ 9.5 \\
\enddata

\end{deluxetable*}

\section{Temperature and Energies} \label{sec:temperature}

\begin{figure*}[t]
\centering
\includegraphics[width=1\textwidth]{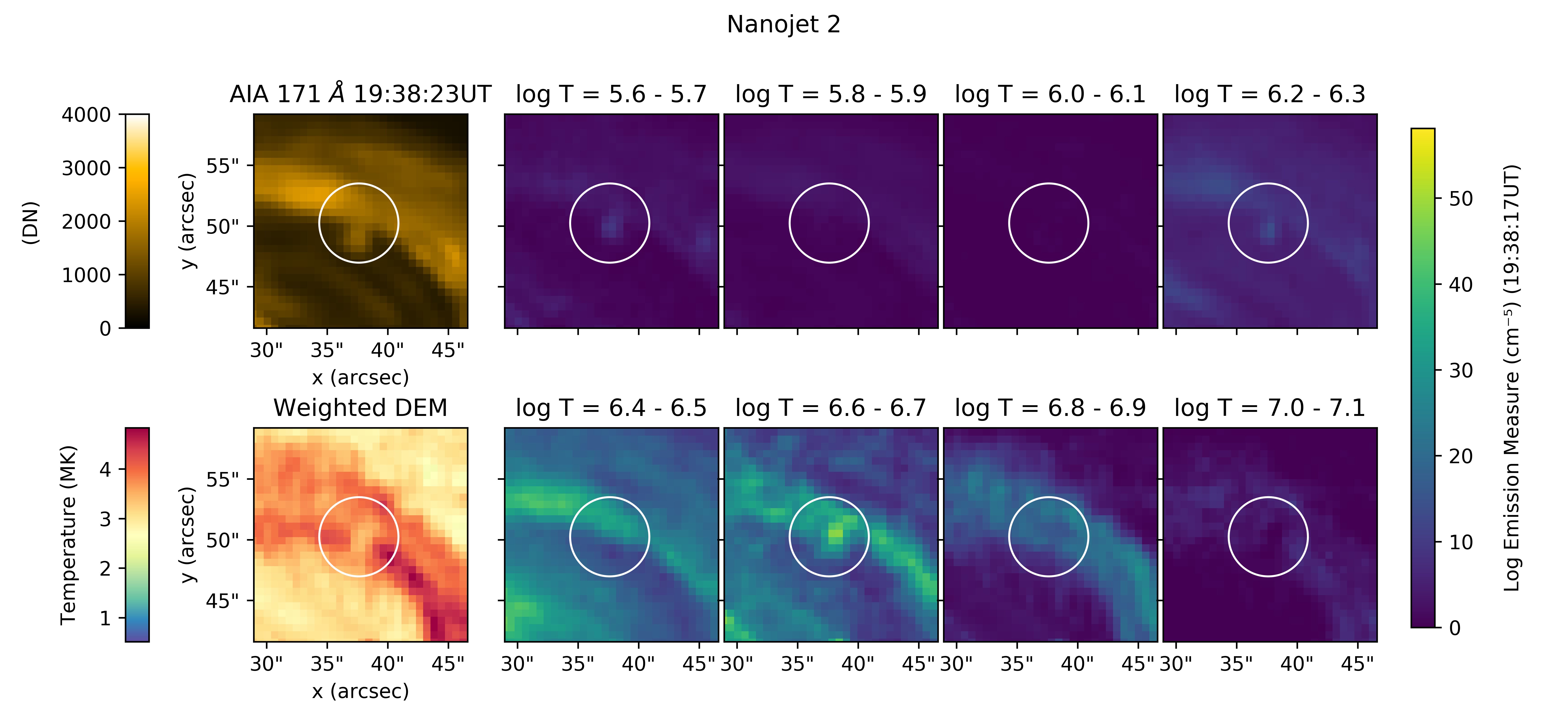}
\caption{The AIA 171 image (top left) for the nanojet in Fig 2 from the loop-like structure above the blowout jet, the weighted DEM temperature of the region plotted in the AIA image (bottom left), and the temperature bins for the emission from the DEM analysis.
\label{fig11}}
\end{figure*}

We have used the Basis Pursuit Method from the Differential Emission Measure (DEM) analysis (\citealt{cheung2015}) to estimate the temperature and densities of the nanojets and the structures where they are found. The nanojets in the blowout jet structure is visible across different AIA channels (see Figure \ref{fig5}), which results in a DEM analysis that can distinguish the structure and nanojets from its surroundings. The visibility of the nanojets in a wide range of wavelength channels also allows us to see the nanojet structure in different temperature bins, as shown in Figure \ref{fig11}. Here, we observe emission from the nanojets in the temperature bins 5.5, 5.6, 6.2, and 6.6, where the head of the nanojet is mostly seen in the 6.2 bin, and the tail is on the lower temperature bins 5.5 and 5.6. The visibility of the nanojets across different IRIS and AIA channels (along with the emission from the different DEM bins) indicates that the nanojets are multithermal. Specifically for the blowout jet where the signatures are clearer, the weighted DEM temperature shows the nanojets as cold spots in comparison to its hotter surroundings, which is a result of this multithermal nature. This is shown in the Weighted DEM plot of Figure \ref{fig11}, as we can see a cold spot in the location that corresponds to the nanojet seen in the AIA 171 channel. Note, however, that the location where the nanojets originate from (as well as the nanojet tail) shows very hot temperatures of log T = 6.6-6.7. This indicates the multithermal nature of the nanojets, as the nanojet contains different temperatures at different parts. Overall, the average temperature for all the nanojets found in the blowout jet is 3.5 MK, which is reasonable considering that most of the emission comes from the log T = 6.6-6.7 bin.

In the other two datasets which contained loops on the limb, we observed dark regions which indicate absorption, and therefore we can also use EUV absorption (\citealt{landi2013}) along with the DEM analysis to measure the number densities of the rain. The dark regions that we have selected match the location of the loop-like structures visible in the SJI 1400, and in particular, are only seen to form in 171 after the cool structure has formed. Prior and after the appearance of the cool structure, these areas are brighter and have uniform intensity with its surroundings, indicating causality with EUV absorption. The EUV absorption method uses Hydrogen and Helium's absorption properties for an absorbing plasma along the line of sight, allowing us to calculate the total number density of the plasma which is a different quantity to what we measured through the DEM analysis (the DEM analysis only provides the electron number density). We have also used the assumption that the thickness of the loop structure along the line of sight is roughly similar to the width of the tube-like structure making the loop. For the loop in the E-limb (after a C class flare), the EUV absorption and DEM analysis suggest a total number density and electron number density values of around $0.9\times10^{10}$~cm$^{-3}$ and $0.4\times10^{10}$~cm$^{-3}$ respectively for the coronal rain. The temperature of the rain measured by the DEM is also around 2.2-2.5 MK. As for the loop-like structure in the W-limb, the EUV absorption and DEM analysis indicate a density of around $5.4\times10^{10}$~cm$^{-3}$ and $1.4\times10^{10}$~cm$^{-3}$ each for the coronal rain. DEM analysis also shows an average temperature of around 2.5 MK for the rain. However, these values may not be representative due to how the rain in both observations is not clearly visible in the AIA channels except for AIA 304, resulting in an emission value that more closely resembles the surrounding corona. The number density value from the DEM analysis, while within the range of commonly observed rain densities, may have been contaminated from the surrounding ambient plasma and thus probably constitutes a lower bound. Therefore, the DEM results from the EUV absorption and DEM analysis may not be reliable for the structure's coronal rain in these observations.

\begin{figure*}[ht]
\centering
\includegraphics[width=1\textwidth]{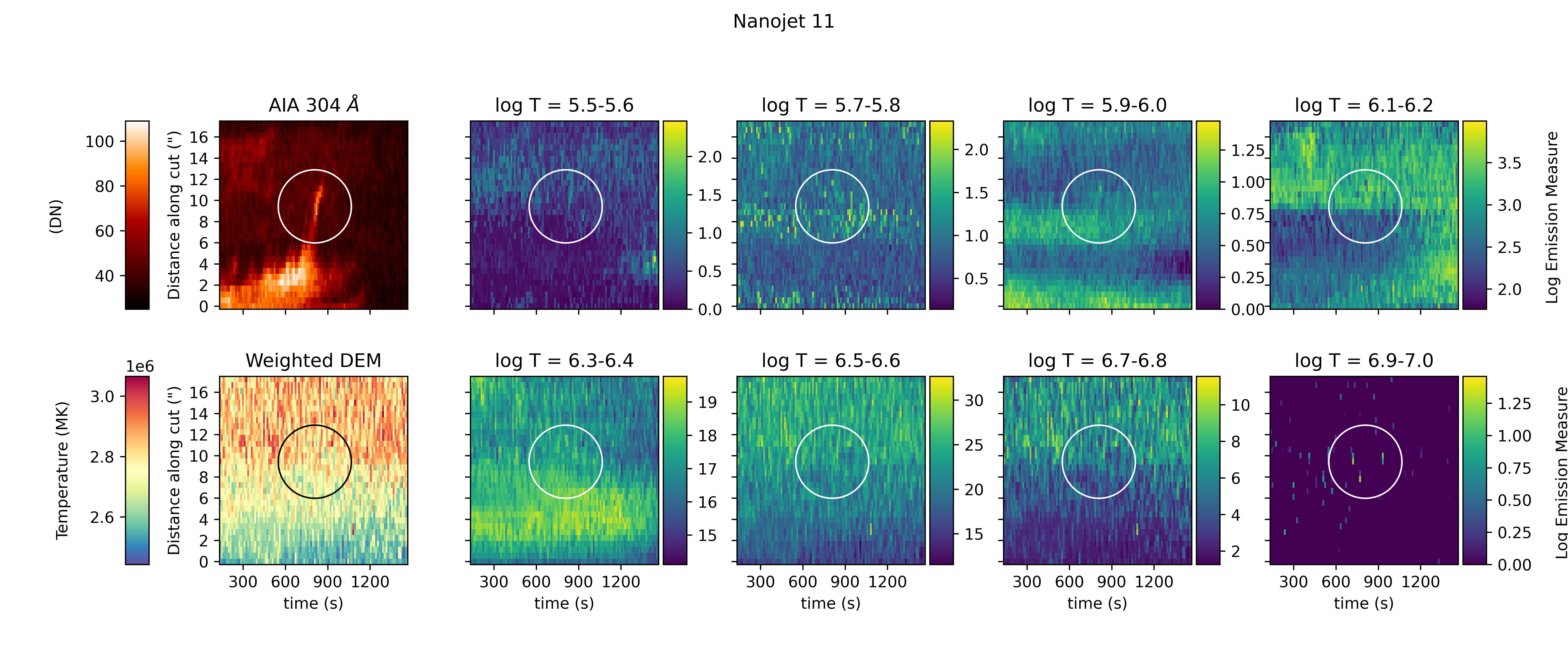}
\caption{The time distance plot from the AIA 304 images for the nanojet in Figure \ref{fig3} from the loop on the E-limb, the weighted DEM temperature of the time distance plot (bottom left), and the temperature bins for the emission from the DEM analysis.
\label{fig12}}
\end{figure*}

\begin{figure*}[hb]
\centering
\includegraphics[width=1\textwidth]{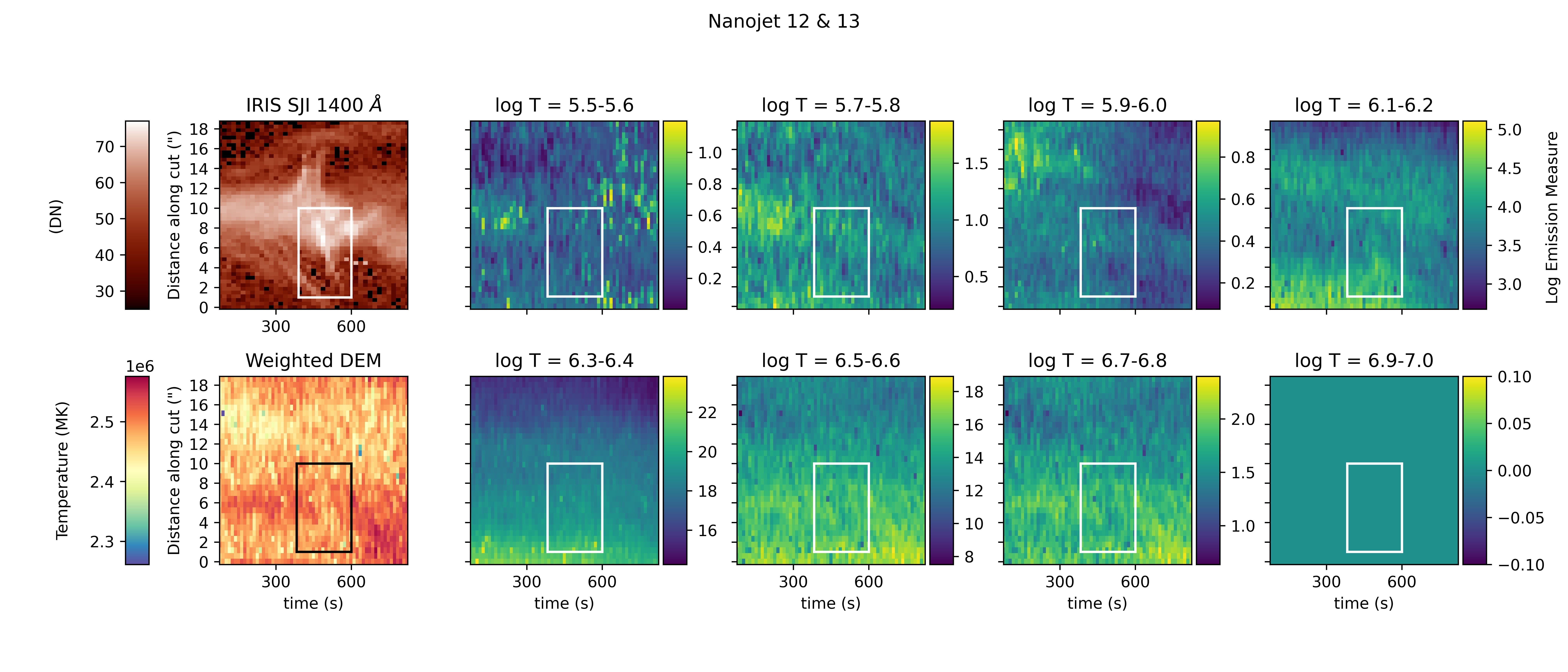}
\caption{
The time distance plot from the IRIS images for the nanojet in Figure \ref{fig4} from the loop on the W-limb, the weighted DEM temperature of the time distance plot (bottom left), and the temperature bins for the emission from the DEM analysis. The rectangle plotted in the bins shows the location where nanojet N12 forms in the time distance diagrams.
\label{fig13}}
\end{figure*}

\begin{figure*}[ht]
\centering
\includegraphics[width=0.95\textwidth]{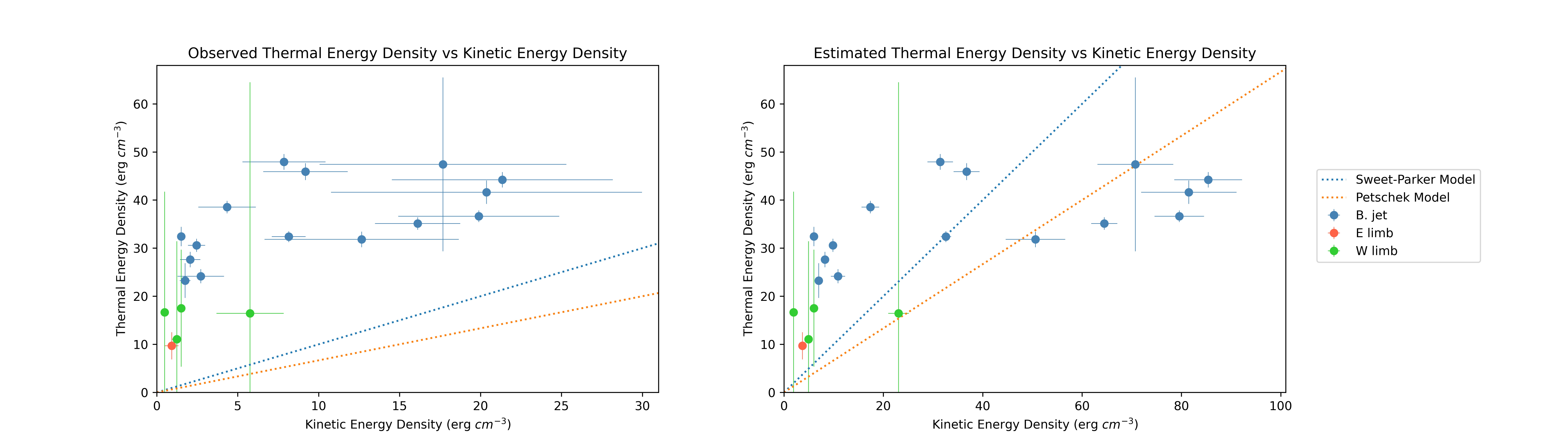}
\caption{
A scatter plot of the observed (left) and estimated (right) kinetic energy density vs thermal energy density of the nanojets, along with the Sweet-Parker and Petschek model.
\label{fig14}}
\end{figure*}

The average temperature of the nanojets is 2.5 MK for all the nanojets found in the two loops, but due to the faint signatures, the DEM analysis only shows faint signals in the individual DEM bins. Snapshots of the DEM images such as in Figure \ref{fig11} did not show any nanojet signatures from these two datasets, and the weighted temperatures from the DEM analysis were not able to clearly distinguish the structure of the rain with its surrounding, unlike in the blowout jet case. However, the signatures are still faintly visible in the DEM images of the time-distance plots from Figure \ref{fig6}, shown in Figures \ref{fig12} and \ref{fig13}. While the loop-like structures are not clear in the AIA channels, the nanojets are still visible in the time-distance diagrams of AIA 171 and 193 as spikes in the diagram (signature also highlighted in a box for Figures \ref{fig13} and \ref{fig14}), and there is faint signature from the DEM log T = 6.1-6.2 bin. While the DEM results from the EUV absorption and DEM analysis may not be the most reliable for these observations, the temperature of around 2.5 MK is not unexpected.

To calculate the properties of the nanojets, we have made the assumption that the nanojets are cylindrical where the height of the nanojet is the length of the cylinder and the width is its diameter. Overall, the average electron number density ($n_{e}$) from the emission calculated from the DEM analysis is $1.9 \pm 0.1 \times 10^{10}$~cm$^{-3}$, which is within the typical coronal rain range \citep{antolin2022}. We also assume that the image pixels that correspond to a nanojet is dominated by the emission from the nanojet itself. Using the measured width and lengths, we then estimate the average kinetic and thermal energies (KE and TE) of the nanojets through the kinetic density ($ E_{K}$) and thermal energy density ($E_{T}$), with $ E_{K} = \frac{1}{2}1.4n_{e}m_{p}v^{2} $ and $E_{T} = \frac{3}{2}2.3n_{e}kT $, where $m_{p}$ is the proton mass, $k$ is the Boltzmann constant, and $T$ is the temperature. The factor of 2.3 in $E_T$ comes from the assumption that the plasma is mostly composed of Hydrogen and Helium, and almost fully ionised. This gives us the average kinetic and thermal energy values of $2.8\pm1.6\times10^{24}$ erg and $1.7\pm1.0\times10^{25}$ erg respectively; which is within the nanoflare range. It is worth noting that the value for the kinetic energy is likely to be underestimated, as we have only used the velocity obtained through the plane of sky. The structures that we have observed appear to be at an angle from the plane of sky, meaning that the actual velocity of the nanojets would also have a Doppler velocity component. We can expect the Doppler velocity on the same order of magnitude as the POS component. This means roughly a factor of 4 larger kinetic energies on average. 

We also compared the kinetic and thermal energy densities distribution to the Sweet-Parker and Petschek models in Figure \ref{fig14}, and found that the distribution differs from what is expected from the two models. In the Sweet-Parker model, the magnetic energy is converted into kinetic and thermal energy at equal amounts, whereas for the Petschek model, 3/5 of magnetic energy is converted into kinetic energy and the remaining converted into thermal energy \citep{priest2014}. This discrepancy could be attributed to the fact that Doppler velocities are not included in our velocity calculations. For the sake of argument, assuming that Doppler velocities are of the same order as the POS velocities (taking them as equal), we can recalculate the kinetic energies. We show these new values on the right panel of Figure~\ref{fig14}. As can be seen, the energies of the nanojets would shift closer to the two lines representing the Sweet-Parker and Petschek models.

\section{Discussion} \label{sec:discussion}

The observations of the nanojets reported in this paper broadly agrees with the nanojets found by \cite{antolin2020} and supports the theory and interpretation that nanojets are a product of episodic small-scale reconnections from field misalignments. We observe equal amounts of nanojets that occur as a single event as in clusters, with 10 for each. From all of the counts, nearly all of the nanojets occur in regions close to the apex and on the edges of the structures, with a few exceptions where the nanojet appears not to be in the structure's edges. However, in these cases we cannot determine whether they indeed occur within the structure or not, due to the line of sight that causes the overlap of the nanojets with the structures themselves. It must also be noted that the numbers found here is non-exhaustive, but just a representative list based on the most visible ones in our observations. There are other instances in the datasets where we have identified nanojet-like structures, but they are too small to provide clear signatures that can be used for analysis.

From the right scatter plot of Figure \ref{fig14}, a noteworthy factor is the distribution of the nanojets into two separated groups with different trends. The nanojets with the larger kinetic and thermal energies tend to follow the Petschek model, while the nanojets with the smaller kinetic and thermal energies are closer to the Sweet-Parker model. Although we do not have a statistically meaningful number of nanojets and the Doppler velocities have been assumed to be of the same order as the POS velocities, this energy distribution could be attributed to a self-regulating mechanism of energy release (\citealt{uzdensky2007}), where under-dense and collisionless plasma correspond to the Petschek model and the dense plasma correspond to the Sweet-Parker model. \cite{uzdensky2007} suggested that the plasma may follow a periodic cycle where it switches between the two models: A low density coronal loop with accumulated free energy will experience fast reconnection following the Petschek model, leading to large amounts of energy release and therefore chromospheric evaporation. This, in turn leads to a density increase, which makes the reconnection collisional and slow, following the Sweet-Parker model. This allows the free energy to build again, and over time the loop depletes, going back to the first step (\citealt{imada2012}). In our context, the blowout jet may be occuring in a loop with initially large magnetic free energy, which is supported by the evidence of shear flows. A fast Petscheck-type reconnection is therefore expected initially, but due to the large increase of density produced by the blowout jet we would expect a rapid transition to the slow collisional Sweet-Parker-type reconnection. On the other hand, the coronal rain cases seem to correspond to Sweet-Parker reconnection, which is explained by the dense coronal rain. A larger statistical investigation is needed to confirm this hypothesis.

When compared with the nanojets in \cite{antolin2020}, our average length for the nanojets are also longer, but the average widths are similar. The driver for the nanojets in this paper would also differ as we do not have a prominence nearby that destabilises and acts as a catalyst for the reconnections. Another interesting peculiarity of the nanojets is their unidirectional nature, in apparent contradiction with the reconnection picture. Indeed, bi-directional jets are expected from reconnection, and it has been conjectured that the loop's curvature or braiding may be the cause for this unique trait \citep{pagano2021}. This is seen in the W-limb loop where we have observed two bi-directional jets, N12 and N13, which demonstrates the opposite direction of travel. While we have yet to see nanojets within a cluster where the nanojets are bi-directional, we still suspect that this is a possibility.

According to the theoretical interpretation of nanojets, nanojets correspond mainly to the sideways field line motion that follows magnetic reconnection at small angles in a guiding field (here, the coronal loop). Hence, there is no major bulk flow as such (as also demonstrated by the simulations in \citealt{antolin2020}) with any resulting (secondary) field-aligned flow only a small fraction of the total speed (10\% or so). If the magnetic field strength is stronger towards the loop axis then the sideways motion directed radially inwards should be smaller than that directed outwards. However, we would expect the field strength difference to be very small and it is unclear if it would be enough to explain the observed unidirectional nature of nanojets.

In terms of coronal heating, nanojets may indeed help in sustaining the structure at coronal temperatures, as may be the case for the blowout jet.  We can approximate the radiative cooling time as $t_{rad} = 30(3\times10^{10}/n)(T/10^6)$~s, where n denotes the average number density and T the average temperature of the cooling loop (e.g. Eq. 18 in \citealt{antolin2022}). This approximation is obtained by taking an average value of $\Lambda=10^{-21.31}$ and $\alpha=0$ for the optically thin loss function (based on \citealt{colgan2008}). In the case of the blowout jet, after its occurrence we observe values of log T = 6.5 and $n=10^{10}$~cm$^{-3}$ along the loop. We obtain average radiative cooling times of ~5 min. On the other hand, the loop is seen to live for 20 min, out of which the nanojets are seen during 6 minutes. This therefore suggests that the nanojets play a role in the maintenance of the loop at coronal temperatures. The exact contribution would depend on their numbers. For the blowout jet we clearly observe 15 nanojets but very likely many more are present as part of clusters or at the limit of the IRIS resolution (as visual inspection suggests). A rough order of magnitude estimate indicates that we would need about 100-150 events to fully sustain the coronal loop (as in the case of \citealt{antolin2020}). On the other hand, for the other two cases where nanojets are seen in coronal rain, we only observe few nanojets with minor coronal heating, and the observed loop structures do not reheat to coronal temperatures following the nanojet occurrence. However, these may play a role in slowing down the catastrophic cooling that accompanies the formation of coronal rain.

\subsection{The role of KHI in nanojet generation}

Here, we also hypothesise that the cluster occurrence of nanojets may indicate the existence of multiple misalignments in the local field, and possibly the existence of twist in the structure. In this scenario, instabilities can help drive the reconnection between misaligned field lines (or produce misalignments themselves). We investigate this possibility by first considering the KHI. The critical velocity for the KHI to develop can also be derived from \cite{hasegawa1975}
\begin{equation}
(\vec{k}\cdot\vec{v_{1}} - \vec{k}\cdot\vec{v_{2}})^2 > \frac{(\rho_{1} + \rho_{2})}{\mu_{0}\rho_{1}\rho_{2}}[(\vec{k}\cdot\vec{B_{1}})^2 + (\vec{k}\cdot\vec{B_{2}})^2]
\end{equation}
where $\vec{v_{1}}$ is the velocity inside the structure, $\vec{v_{2}}$ is the surrounding velocity outside the structure, $\vec{k}$ is the wave vector, $\vec{B_{1}}$ is the magnetic field inside the structure, $\vec{B_{2}}$ is the magnetic field outside the structure, $\rho_1$ and $\rho_2$ are the number densities inside and outside the structure, and $\mu_{0}$ is the magnetic permeability. Assuming that the flow introduces the perturbations, we take $\vec{k}||\vec{v_{1}}$ and that $\vec{k}$ and $\vec{v_{2}}$ are at an angle $\theta$, the equation can be reduced to obtain
\begin{equation} \label{eq:2}
    v_{1} - v_{2}\cos(\theta) > \sqrt{\frac{\rho_{1}+\rho_{2}}{\mu_{0}\rho_{1}\rho_{2}} (B_{1}^2 + B_{2}^2 \cos^2(\theta))}
\end{equation}
Additionally, we can also assume that $B_{1}\approx B_{2}$ and $v_{1} >> v_{2}$, and that the density inside the loops satisfy $\rho_{1} >> \rho_{2}$, which gives us the critical velocity inside the loop $v_{1,crit}$ to be
\begin{equation} \label{eq:3}
    v_{1,crit} = v_{A,2}\sqrt{1 + \cos^2(\theta)}
\end{equation}
where $v_{A,2} = \frac{B_2}{\sqrt{\mu_0\rho_2}}$, which is the Alfv\'en speed of the surrounding medium.

\begin{figure*}[ht]
\centering
\includegraphics[width=0.8\textwidth]{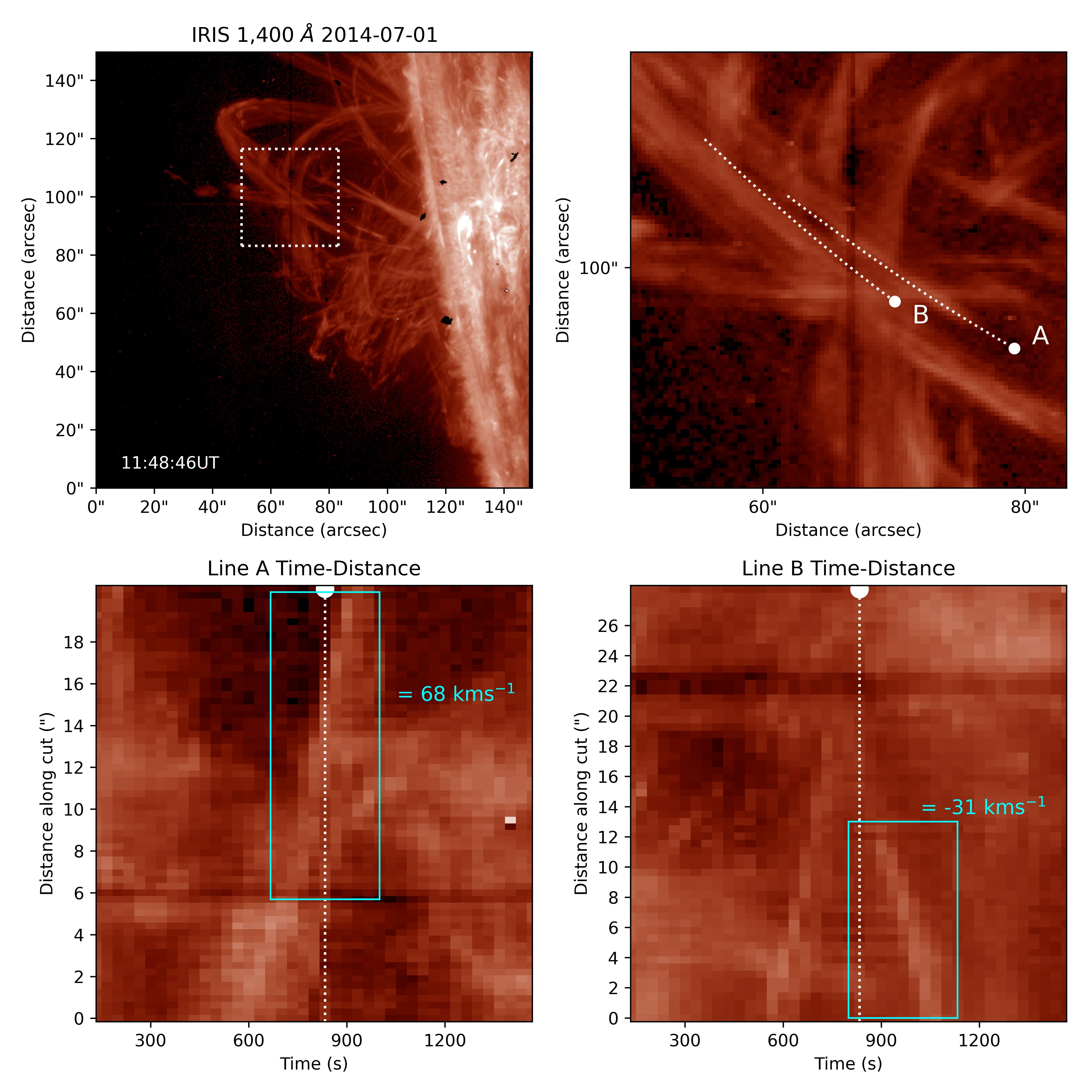}
\caption{
Counter-streaming flows identified in the E-limb loop. Top row (left to right) shows a snapshot of the structure showing the location of one of the loop's legs, where we have taken 2 rain clump paths A and B for time-distance diagrams shown in the right figure. The time-distance diagrams are plotted on the bottom row, with the cyan boxes indicating the flows that we have used for the velocity measurement and the corresponding POS velocities. The white dotted line at the center of the time-distance plots indicates the time of the snapshots from the top row. Note the opposite slopes indicating the counter-streaming flows.
\label{fig15}}
\end{figure*}

\cite{li2018} reported the occurrence of the KHI in the coronal loop due to the blowout jet, where there are upflow velocities that reach up to 224-476~km~s$^{-1}$ and a shear velocity of 204~km~s$^{-1}$ between two flows inside the loop-like structure produced by the jet. This also prompts the question of whether there is shear velocity within the structure for the other two datasets, or between the structure and the surrounding material which we assume to be plasma at rest. The slit of \textit{IRIS} crosses through the two footpoints of the loop for datasets 2 and 3, allowing us to measure the observed rest wavelength of Mg II k. We find 2796 \r{A} as the rest wavelength for the 2 datasets, with the assumption that the angle between the LOS and the loop plane is the same at both leg crossings. For the E-limb loop, we observe a maximum Doppler speed of 52~km~s$^{-1}$ from one of the rain clumps travelling downwards at 68~km~s$^{-1}$ in the POS, which would give a a true velocity of 86~km~s$^{-1}$. We also observed a different rain clump nearby travelling at -31~km~s$^{-1}$ (ie. along the opposite direction) in the POS, which gives us a true velocity value of -61~km~s$^{-1}$, and therefore, an internal shear of 147~km~s$^{-1}$. The time-distance diagrams showing the flows observed used for these measurements are shown in Figure \ref{fig15}. Whereas for the W-limb loop, the largest Doppler speed observed is only 9~km~s$^{-1}$ from a flow that travels downwards with a speed of 137~km~s$^{-1}$ in the POS, meaning that its true velocity is 138~km~s$^{-1}$. A nearby flow also travels in the same direction with a velocity of 70~km~s$^{-1}$, meaning that the internal shear velocity is 68~km~s$^{-1}$.

While these are the values that we observe from the material that crosses the slit, we suspect that there could still be larger Doppler speeds at other parts of the structure which may result in a larger shear. This could indicate that the presence of shear velocity could promote the growth of KHI and therefore be a catalyst to producing nanojets. We therefore calculate the $v_{crit}$ for the KHI to form from the shear between the flows in or surrounding the structure using different parameters, which are given in Table \ref{table2}. From the table we find that $v_{crit}$ decreases as $\theta$ and $n_2$ increases, but $v_{crit}$ increases as $B_2$ becomes larger. For the lower number densities representing the surrounding plasma ($n_2 = 10^8 - 10^9$~cm$^{-3}$), $v_{crit}$ is always larger than 356~km~s$^{-1}$, which is larger than the velocity of the flows observed in the east and west limb loops. This suggests that the KHI cannot occur between the shear flow from the structure and the surrounding environment in our observations, but as $v_{crit}$ is lower than 130~km~s$^{-1}$ for $n_2 = 10^{10}$~cm$^{-3}$, this means that the KHI may still form from the observed internal shear between two flows. Indeed, for the E-limb loop, the fastest shear flow observed is 147~km~s$^{-1}$. This value is larger than the $v_{crit}$ values for $n_2 = 10^{10}$~cm$^{-3}$ and $B = 5$~G in Table \ref{table2}, and for $n_2 = 10^{11}$~cm$^{-3}$ when  $B < 20$~G, meaning that the KHI may form in weak magnetic fields. Coronal rain densities are usually on the order of $10^{10} - 10^{11}$~cm$^{-3}$ (\citealt{antolin2022}), which is supported by our DEM analysis. On the other hand, for W-limb loop, the internal shear of 68~km~s$^{-1}$ that we observe is smaller than the calculated $v_{crit}$ in Table \ref{table2} for $n_2 \leq 10^{10}$~cm$^{-3}$, meaning that the KHI is less likely to form. It is only when $n_2 = 10^{11}$~cm$^{-3}$ when the observed shear is larger than three of the calculated $v_{crit}$, which is when $B_2 = 5$~G.

\begin{figure}[hb]
  \centering
  \includegraphics[width=0.3\textwidth]{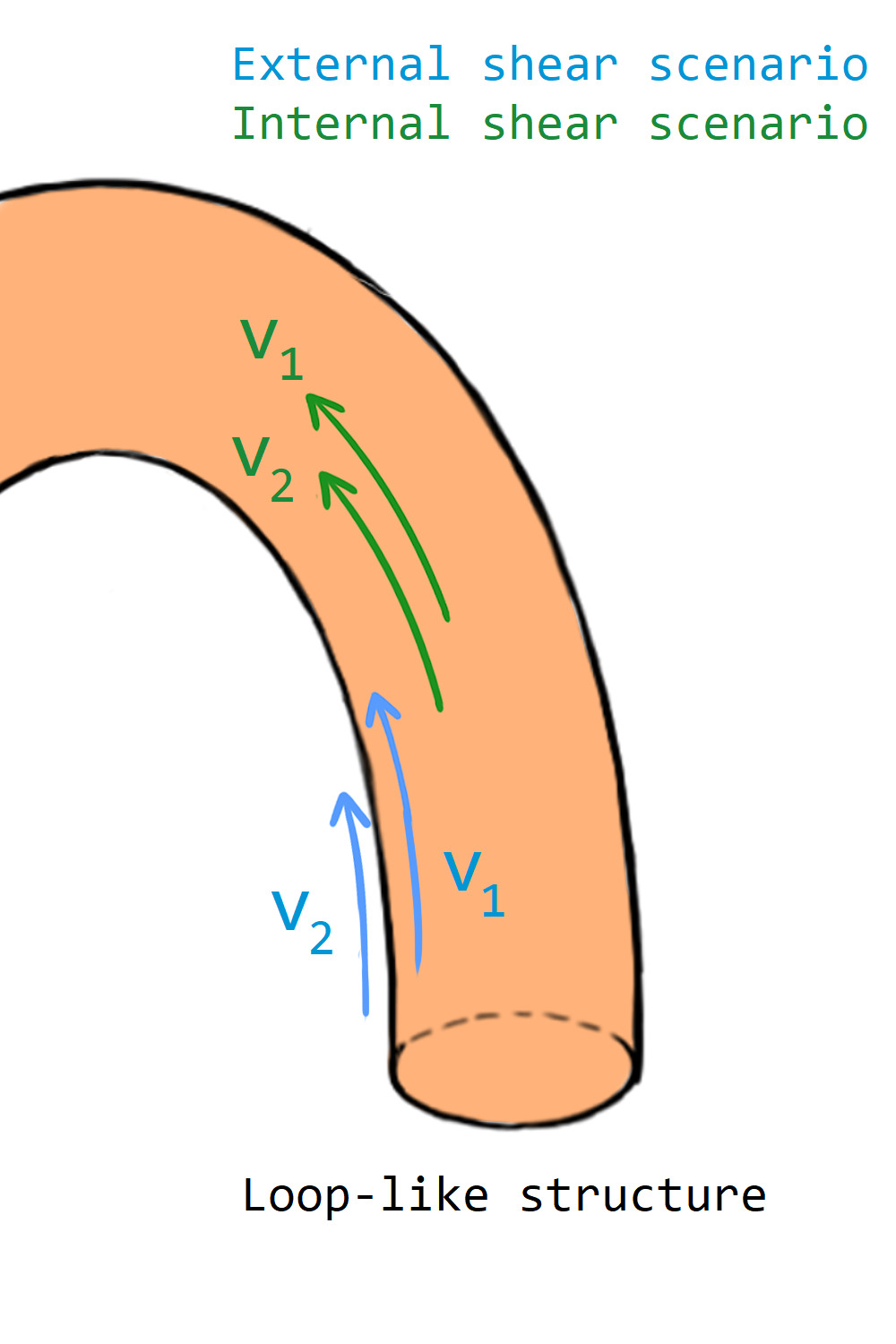}
  \caption{A schematic showing the internal and external shear scenarios in Table \ref{table2}.
  \label{fig16}}
\end{figure}

\begin{deluxetable*}{@{\extracolsep{7pt}}cccccc|cccccc}
\tablenum{2}
\tablecaption{The calculated critical velocity ($v_{crit}$) required for the KHI to occur using Equation \ref{eq:3}, using the angle between $\vec{k}$ and $\vec{v_{2}}$, surrounding number density $n_{2}$, surrounding magnetic field $B_{2}$, and Alfv\'en speed $v_{A,2}$. The column KHI states for which dataset the KHI is possible with the given parameters (no entry means that the given parameters cannot induce the KHI for any of the datasets). The values that differ for the external and internal shear scenario are the $n_2$, where the internal shear flow has larger number densities. The two scenarios (external and internal shear) is described in Figure \ref{fig16}.}\label{table2}
\tablewidth{0pt}
\tablehead{
  \multicolumn{5}{c}{External Shear Scenario}&
  \multicolumn{6}{c}{Internal Shear Scenario}&\\
\cline{1-6} \cline{7-12}
\colhead{$\theta$} & \colhead{$n_{2}$} & \colhead{$B_{2}$} & \colhead{$v_{A,2}$}& \colhead{$v_{crit}$} & \colhead{KHI} & \colhead{$\theta$} & \colhead{$n_{2}$} & \colhead{$B_{2}$} & \colhead{$v_{A,2}$}& \colhead{$v_{crit}$} & \colhead{KHI}\\
\colhead{($^{\circ}$)} & \colhead{($cm^{-3}$)} & \colhead{(G)} & \colhead{(~km~s$^{-1}$)}&  \colhead{(~km~s$^{-1}$)} & \colhead{(data)}& \colhead{($^{\circ}$)} & \colhead{($cm^{-3}$)} & \colhead{(G)} & \colhead{(~km~s$^{-1}$)}&  \colhead{(~km~s$^{-1}$)} & \colhead{(data)}
}
\startdata
   5 & $10^{8}$ & 5 & 921 & 1300 &    & 5 & $10^{10}$ & 5 & 92 & 130 & B jet and E limb \\
   5 & $10^{9}$ & 5 & 291 & 411 &    & 5 & $10^{11}$ & 5 & 29 & 41 & all \\
   5 & $10^{8}$ & 10 & 1842 & 2600 &    & 5 & $10^{10}$ & 10 & 184 & 260 &  \\
   5 & $10^{9}$ & 10 & 583 & 822 &    & 5 & $10^{11}$ & 10 & 58 & 82 & B jet and E limb \\
   5 & $10^{8}$ & 20 & 3685 & 5201 &    & 5 & $10^{10}$ & 20 & 368 & 520 &  \\
   5 & $10^{9}$ & 20 & 1165 & 1644 &    & 5 & $10^{11}$ & 20 & 117 & 164 & B jet \\
   5 & $10^{8}$ & 25 & 4606 & 6501 &    & 5 & $10^{10}$ & 25 & 461 & 650 &  \\
   5 & $10^{9}$ & 25 & 1457 & 2055 &    & 5 & $10^{11}$ & 25 & 146 & 205 &  \\
   25 & $10^{8}$ & 5 & 921 & 1243 &    & 25 & $10^{10}$ & 5 & 92 & 124 & B jet and E limb \\
   25 & $10^{9}$ & 5 & 291 & 393 &    & 25 & $10^{11}$ & 5 & 29 & 39 & all \\
   25 & $10^{8}$ & 10 & 1842 & 2486 &    & 25 & $10^{10}$ & 10 & 184 & 248 &  \\
   25 & $10^{9}$ & 10 & 583 & 786 &    & 25 & $10^{11}$ & 10 & 58 & 78 & B jet and E limb \\
   25 & $10^{8}$ & 20 & 3685 & 4973 &    & 25 & $10^{10}$ & 20 & 368 & 497 &  \\
   25 & $10^{9}$ & 20 & 1165 & 1572 &    & 25 & $10^{11}$ & 20 & 117 & 157 & B jet \\
   25 & $10^{8}$ & 25 & 4606 & 6216 &    & 25 & $10^{10}$ & 25 & 461 & 621 &  \\
   25 & $10^{9}$ & 25 & 1457 & 1965 &    & 25 & $10^{11}$ & 25 & 146 & 196 & B jet \\
   45 & $10^{8}$ & 5 & 921 & 1128 &    & 45 & $10^{10}$ & 5 & 92 & 112 & B jet and E limb \\
   45 & $10^{9}$ & 5 & 291 & 356 &    & 45 & $10^{11}$ & 5 & 29 & 35 & all \\
   45 & $10^{8}$ & 10 & 1842 & 2256 &    & 45 & $10^{10}$ & 10 & 184 & 225 &  \\
   45 & $10^{9}$ & 10 & 583 & 713 &    & 45 & $10^{11}$ & 10 & 58 & 71 & B jet and E limb \\
   45 & $10^{8}$ & 20 & 3685 & 4513 &    & 45 & $10^{10}$ & 20 & 368 & 451 &  \\
   45 & $10^{9}$ & 20 & 1165 & 1427 &    & 45 & $10^{11}$ & 20 & 117 & 142 & B jet and E limb \\
   45 & $10^{8}$ & 25 & 4606 & 5641 &    & 45 & $10^{10}$ & 25 & 461 & 564 &  \\
   45 & $10^{9}$ & 25 & 1457 & 1783 &    & 45 & $10^{11}$ & 25 & 146 & 178 & B jet
\enddata
\end{deluxetable*}

If there is an angle between the wave vector and the magnetic field, then the KHI is more likely to form along the flow due to the shear velocity with its surroundings, since this reduces the magnetic tension opposing the instability growth. This is seen in Table \ref{table2} which shows that larger angles have a smaller $v_{crit}$ to trigger the instability. For the blowout jet, \cite{li2018} noted that there may be magnetic untwisting in the structure with a pitch angle of 30-60$^{\circ}$, whereas for the loops in the E-limb and loop like structure in the W-limb, we observe angles of 5-10$^{\circ}$.

\subsection{Rayleigh-Taylor Instability}

\begin{figure}[hb]
  \centering
  \includegraphics[width=0.43\textwidth]{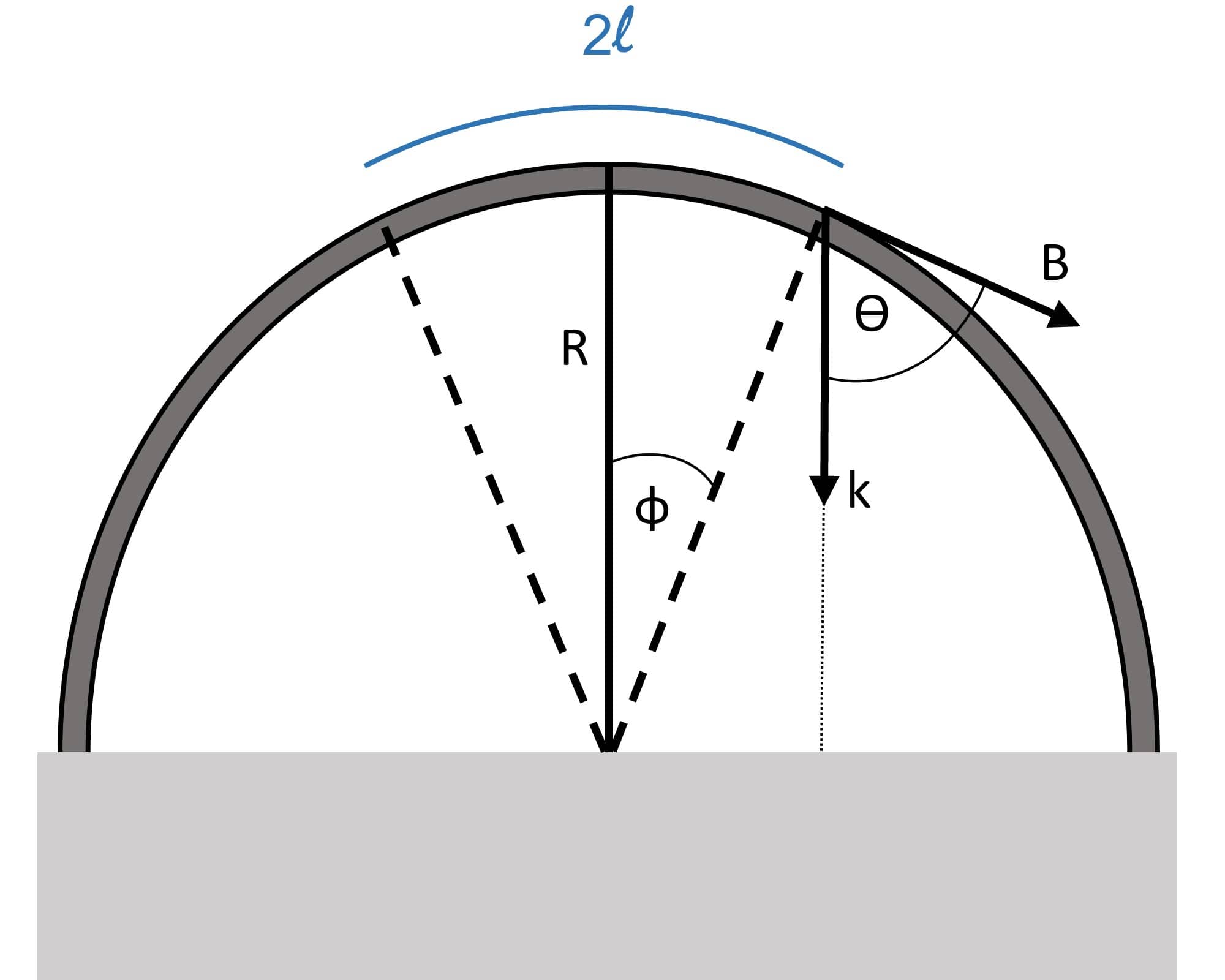}
  \caption{A schematic showing the angles in a loop from equation (4) and (5) indicating a region l where the Rayleigh Taylor Instability is likely to occur.
  \label{fig17}}
\end{figure}

A common feature between all observed flows are the large densities. Since these occur near the apex of the loops, another possible instability is the Rayleigh-Taylor Instability (RTI). We investigate here the possibility of the RTI being the trigger of the nanojets. We can calculate the growth rate of the RTI using the following equation from \cite{chandrasekhar1961}:

\begin{equation}
    \sigma^2 = gAk - \frac{(\vec{k}\cdot\vec{B})^2}{2\pi(\rho_{1}-\rho_{2})} 
\end{equation}
where
\begin{equation}
    A = \frac{\rho_{1}-\rho_{2}}{\rho_{1}+\rho_{2}}
\end{equation}
is the Atwood number, $\rho_{1}$ is the density within the loop, $\rho_{2}$ is the surrounding density, g is the gravity, $k=\frac{2\pi}{\lambda}$ is the wavenumber, and B is the magnetic field strength. Taking $\theta$ as the angle between the wave vector and the magnetic field, the minimum angle $\theta_{min}$ required for the RTI to form can also be derived from the above equations to obtain 
\begin{equation}
    \cos(\theta_{min})=\frac{\sqrt{gAk2\pi(\rho_{1}-\rho_{2})}}{kB}
\end{equation}
The minimum angle above also determines the region where the RTI is likely to form based on Figure \ref{fig17}, and we can calculate the distance $l$ over which the rain plasma is subject to similar favourable geometric conditions for the RTI to occur.
\begin{equation}
    \phi = \frac{\pi}{2} - \theta
\end{equation}
\begin{equation}
    l = R\phi
\end{equation}
\begin{equation}
    dt = \frac{2l}{v}
\end{equation}
where $R$ is the height of the loop, $v$ is the observed velocity of the rain at the apex and dt corresponds to the time the rain is within the region of length $l$. 

Using coronal values, $R = 34,000$~km, and the assumption that $\lambda=600$~km (which is close to the average width of a nanojet), and a velocity similar to the projected velocity at the apex of the structures of $25$~km~s$^{-1}$, we calculated the minimum growth time $1/\sigma_{min}$ (ie. when $\theta = 90^{\circ}$) it takes for the RTI to form. We then find a region within $2l$ where the growth time is less than $dt$, and therefore calculate the average of the growth time in that region $1/\sigma_{eff}$, or the 'effective growth time', and the effective growth rate $\sigma_{eff}$. $1/\sigma_{eff}$ is only calculated if $1/\sigma_{min}$ is smaller than $dt$, as it means that the plasma clump that travels through the arclength $l$ will have enough time to allow the onset of the RTI. As this new region is smaller and have a shorter travel time, we then calculated the ratio $\tau$ between this new travel time $dt_{eff}$ and the $1/\sigma_{eff}$ as a measure of how likely the RTI can form in the structure. A value above/below 1 indicates that the RTI is likely/unlikely to occur, respectively. Table \ref{table3} calculates these properties of the RTI, including the effective growth time $1/\sigma_{eff}$, its corresponding effective growth rate $\sigma_{eff}$, and the ratio $\tau$, using different parameters (further breakdown of values for the different observations are shown in Table \ref{table3})

From Table \ref{table3}, we see that 3 sets of parameters have $dt$ less than $1/\sigma_{min}$, meaning that the RTI is unlikely to occur for these cases. For the other instances, the ratio $\tau$ ranges between 0.363 to 1.979. The ratio increases as the difference between $n_1$ and $n_2$ becomes larger, or if the magnetic field decreases. For the typical coronal rain values ($n_1 = 10^{10} - 10^{11}$~cm$^{-3}$), we see that $\tau$ is always over 1 for B = 10~G, but only over 1 for B = 20~G when $n_1 \geq 10^{11}$~cm$^{-3}$. $\tau$ is always exceptionally large for $n_1 > 10^{11}$~cm$^{-3}$. This means that the RTI is likely to form under usual coronal rain conditions, but only for weak magnetic fields that may not be expected in active regions (\citealt{nakariakov2001}; \citealt{kriginsky2021}). This aligns with the work by \cite{moschou2015} and \cite{xia2017}, who have simulated coronal condensation which shows the formation of RTI using weak magnetic fields of less than 10~G at the apex of the loop. However, it is unlikely that RTI is producing the nanojets in the E-limb loop because these are oriented sideways, and not downwards as expected from RTI. On the other hand, two of the nanojets in the W-limb loop do seem to be vertically oriented. This, combined with the theoretical estimations suggest that RTI is the reconnection driver for the W-limb observation.

Plasma beta should be small as these structures are in the corona and have a guide field, where the overall magnetic field topology would only have a small variability throughout the event. If the plasma beta is close to 1, then we would observe some deformation in the structure. We do consider this possibility (but still with a plasma beta $< \sim{}$1) for the case of the loop at the W-limb, where our theoretical analysis indicates that the Rayleigh-Taylor instability is possible. Such scenario is also backed-up by numerical simulations by \cite{moschou2015} and \cite{xia2017}. However, it is important to note that contrary to these simulations where RT leads to coronal rain clearly falling vertically prior to following the field lines, the coronal rain that we observe in all cases after the nanojet event always follow the field lines without any observable deformation of the loop structure. The deformation may still happen at lower spatial scales below the IRIS spatial resolution.

\begin{deluxetable*}{cccccccccccc}
\tablenum{3}
\tablecaption{Growth rate of the Rayleigh Taylor Instability $\sigma$ calculated through equation (4). The other properties used for the calculations and the variables calculated are the total number densities inside and outside of the loop ($n_{1}$ and $n_{2}$), magnetic field (B), Atwood number (A), the minimum angle between $\vec{k}$ and $\vec{B}$ required for the instability to occur ($\theta_{min}$), the distance of the path along which the instability can theoretically occur (l), the time interval in which the rain clump with velocity $v = 25$~km~s$^{-1}$ is within the path l (dt), minimum growth time of the instability ($1/\sigma_{min}$), effective growth time ($1/\sigma_{eff}$), effective growth rate ($\sigma_{eff}$), effective travel time ($dt_{eff}$), and ratio ($\tau$) between $dt_{eff}$ and $1/\sigma_{eff}$.}\label{table3}
\tablewidth{0pt}
\tablehead{
\colhead{$n_{1}$} & \colhead{$n_{2}$} & \colhead{B} & \colhead{A} & \colhead{$\theta_{min}$} & \colhead{l} & \colhead{dt} & \colhead{$(1/\sigma)_{min}$} & \colhead{$(1/\sigma)_{eff}$} &  \colhead{$\sigma_{eff}$} & \colhead{$dt_{eff}$} & \colhead{$\tau$}\\
\colhead{($cm^{-3}$)} & \colhead{($cm^{-3}$)} & \colhead{(G)} & \colhead{} & \colhead{($^{\circ}$)} & \colhead{(km)} & \colhead{(s)} & \colhead{(s)} & \colhead{(s)}  & \colhead{(s$^{-1}$)} & \colhead{s} & \colhead{}}
\startdata
   $10^{10}$ & $10^9$ & 10 & 0.8182 & 89.07 & 551 & 44 & 21 & 32 & 0.023 & 39 & 1.204 \\
   $10^{10}$ & $10^8$ & 10 & 0.9802 & 88.93 & 633 & 51 & 19 & 35 & 0.02 & 47 & 1.352 \\
   $10^{11}$ & $10^8$ & 10 & 0.998 & 86.58 & 2029 & 162 & 19 & 91 & 0.006 & 161 & 1.782 \\
   $10^{12}$ & $10^8$ & 10 & 0.9998 & 79.11 & 6461 & 517 & 19 & 268 & 0.002 & 517 & 1.929 \\
   $10^{13}$ & $10^8$ & 10 & 1.0 & 53.32 & 21769 & 1742 & 19 & 880 & 0.001 & 1741 & 1.979 \\
   $10^{10}$ & $10^9$ & 20 & 0.8182 & 89.54 & 276 & 22 & 21 & 21 & 0.045 & 8 & 0.363 \\
   $10^{10}$ & $10^8$ & 20 & 0.9802 & 89.47 & 316 & 25 & 19 & 22 & 0.04 & 17 & 0.765 \\
   $10^{11}$ & $10^8$ & 20 & 0.998 & 88.29 & 1014 & 81 & 19 & 50 & 0.012 & 79 & 1.582 \\
   $10^{12}$ & $10^8$ & 20 & 0.9998 & 84.58 & 3216 & 257 & 19 & 138 & 0.004 & 257 & 1.86 \\
   $10^{13}$ & $10^8$ & 20 & 1.0 & 72.62 & 10313 & 825 & 19 & 422 & 0.001 & 825 & 1.955 \\
   $10^{10}$ & $10^9$ & 100 & 0.8182 & 89.91 & 55 & 4 & 21 & * & * & * & * \\
   $10^{10}$ & $10^8$ & 100 & 0.9802 & 89.89 & 63 & 5 & 19 & * & * & * & * \\
   $10^{11}$ & $10^8$ & 100 & 0.998 & 89.66 & 203 & 16 & 19 & * & * & * & * \\
   $10^{12}$ & $10^8$ & 100 & 0.9998 & 88.92 & 642 & 51 & 19 & 35 & 0.019 & 48 & 1.367 \\
   $10^{13}$ & $10^8$ & 100 & 1.0 & 86.58 & 2032 & 163 & 19 & 91 & 0.006 & 162 & 1.782 \\
\enddata
\end{deluxetable*}

\section{Conclusion} \label{sec:conclusion}

We have reported the existence of nanojets in 3 different coronal structures observed by \textit{IRIS} and \textit{SDO}, where they share the similar properties and characteristics to the nanojets reported by \cite{antolin2020}. Despite the very different conditions in the loops presented by these observations (blowout jet powered to thermally unstable loops) they all have in common high speed and dense flows. We have therefore investigated the possibility of occurrence of dynamic instabilities (namely, the KHI and RTI) that would drive the reconnection leading to nanojets. For the KHI, we calculated the theoretical critical velocity values in the structure required to trigger the onset of the KHI, and found that the KHI is more likely to happen either within the structure itself, or on the outer edges of the structure but with a large number density. For the blowout jet study, we find that the observed shear velocities satisfy the criterion for the KHI to be triggered at the edge of the structure, in agreement with \cite{li2018}. As reported in that paper, the onset of the KHI is directly confirmed in this case by the observation of vortex-like structures at the edge of the loop. 

The other nanojets found in the loop-like structures still have an unclear origin as we have not directly observed the KHI or the RTI. From our calculations and given that the observed large densities are common to coronal rain, it is very likely that the KHI forms in the structure with large internal counter-streaming flows (as in the E-limb observation), in the case of weak magnetic fields ($< 20$~G). On the other hand, we find that the RTI is also a possibility, assuming that gravity is the source of the perturbation that triggers the instability. This is particularly the case of the W-limb observation, however, this is also limited to the condition of low magnetic fields and large number densities, with the magnetic field as the parameter being the larger contributor to the likelihood of the RTI. However, the KHI also becomes likely for densities over $10^{11}$~cm$^{-3}$, which is higher than what we measure but still possible for quiescent coronal rain. Such number densities (and even larger) are however, common for flare-driven rain (\citealt{scullion2016}). Weak magnetic fields are not expected in active regions (\citealt{nakariakov2001}; \citealt{heinzel2018}; \citealt{kriginsky2021}), however, they may still be expected at the apex of expanding long loops. Additionally, coronal rain may play an important role in the nanojet formation, as suggested by \cite{antolin2020}, because the partially ionized plasma can act as a catalyst for reconnection.

Different to \cite{antolin2020}, dynamic instabilities are identified here as the reconnection driver for the nanojets, supporting the hypothesis that nanojets are largely independent of the way component (small-angle) reconnection is enforced. The variety of environments where we have discovered nanojets also suggests that nanojets are a general result of reconnection that may have been driven by not only field braiding, but also different instabilities. Additionally, It is likely that these structures may also contain numerous misalignments through braiding prior to the onset of an instability that may contribute to the number of nanojets. 

A main challenge presented when observing nanojets comes from their short lifespan. Many of the nanojets that we observed last for no longer than 20~s, with the shortest lasting for no more than 7.2~s. This becomes a challenge to identify nanojets in observations with cadences longer than 10~s, as we may only be able to capture them in 1-2 frames. Combined with their small sizes, this makes nanojet detection even more challenging. Here, we have used cool material from the blowout jet and coronal rain as a high resolution tracer of the nanojets because of the initially cool temperature. The nanojets we detect have also a cool component in addition to their hot component, making them multi-thermal. This aspect may therefore be a bias of the detection process that we used.  

There are still many open questions to answer regarding the nanojet's driver, and this will serve as an avenue for further investigation through both observations and numerical simulations in coronal heating research. These nanojets carry energies within the nanoflare range, further supporting the nanoflare heating theory. With the current spatial and temporal resolution of coronal observations, the identification of nanojets in evolving structures is still a difficult task, however, the characteristic behaviour such as the perpendicular ejection, fast velocities of around 100-200~km~s$^{-1}$, or the field line splitting after the nanojets can be used as useful guide to identify them. This provides a direction for analysis from incoming future observations, which could provide insight into the heating mechanisms that are found in the solar corona.

\begin{acknowledgments}
P. A. acknowledges funding from his STFC Ernest Rutherford Fellowship (No. ST/R004285/2). J.A.M. acknowledges UK Science and Technology Facilities Council (STFC) support from grant ST/T000384/1. IRIS is a NASA small explorer mission developed and operated by LMSAL with mission operations executed at NASA Ames Research Center and major contributions to downlink communications funded by ESA and the Norwegian Space Centre. \textit{SDO} is part of NASA's Living With a Star Program. All data used in this work are publicly available through the websites of the respective solar missions. Lastly, the authors would like to thank the anonymous referee who provided thorough and detailed comments useful for the manuscript.
\end{acknowledgments}


\begin{thebibliography}{}
\expandafter\ifx\csname natexlab\endcsname\relax\def\natexlab#1{#1}\fi
\providecommand{\url}[1]{\href{#1}{#1}}
\providecommand{\dodoi}[1]{doi:~\href{http://doi.org/#1}{\nolinkurl{#1}}}
\providecommand{\doeprint}[1]{\href{http://ascl.net/#1}{\nolinkurl{http://ascl.net/#1}}}
\providecommand{\doarXiv}[1]{\href{https://arxiv.org/abs/#1}{\nolinkurl{https://arxiv.org/abs/#1}}}

\bibitem[{Ajabshirizadeh {et~al.}(2015)Ajabshirizadeh, Ebadi, Vekalati, \&
  Molaverdikhani}]{ajabshirizadeh2015}
Ajabshirizadeh, A., Ebadi, H., Vekalati, R.~E., \& Molaverdikhani, K. 2015,
  Astrophysics and Space Science, 357, \dodoi{10.1007/s10509-015-2277-8}

\bibitem[{{Anfinogentov} {et~al.}(2015){Anfinogentov}, {Nakariakov}, \&
  {Nistic{\`o}}}]{anfinogentov2015}
{Anfinogentov}, S.~A., {Nakariakov}, V.~M., \& {Nistic{\`o}}, G. 2015, \aap,
  583, A136, \dodoi{10.1051/0004-6361/201526195}

\bibitem[{{Antolin} \& {Froment}(2022)}]{antolin2022}
{Antolin}, P., \& {Froment}, C. 2022, FrASS, \dodoi{10.3389/fspas.2022.820116}

\bibitem[{{Antolin} {et~al.}(2015){Antolin}, {Okamoto}, {De Pontieu},
  {Uitenbroek}, {Van Doorsselaere}, \& {Yokoyama}}]{antolin2015}
{Antolin}, P., {Okamoto}, T.~J., {De Pontieu}, B., {et~al.} 2015, \apj, 809,
  72, \dodoi{10.1088/0004-637X/809/1/72}

\bibitem[{{Antolin} {et~al.}(2021){Antolin}, {Pagano}, {Testa}, {Petralia}, \&
  {Reale}}]{antolin2020}
{Antolin}, P., {Pagano}, P., {Testa}, P., {Petralia}, A., \& {Reale}, F. 2021,
  Nature Astronomy, 5, 54, \dodoi{10.1038/s41550-020-1199-8}

\bibitem[{{Antolin} {et~al.}(2018){Antolin}, {Schmit}, {Pereira}, {De Pontieu},
  \& {De Moortel}}]{antolin2018}
{Antolin}, P., {Schmit}, D., {Pereira}, T.~M.~D., {De Pontieu}, B., \& {De
  Moortel}, I. 2018, \apj, 856, 44, \dodoi{10.3847/1538-4357/aab34f}

\bibitem[{{Aschwanden} {et~al.}(2002){Aschwanden}, {de Pontieu}, {Schrijver},
  \& {Title}}]{aschwanden2002}
{Aschwanden}, M.~J., {de Pontieu}, B., {Schrijver}, C.~J., \& {Title}, A.~M.
  2002, \solphys, 206, 99, \dodoi{10.1023/A:1014916701283}

\bibitem[{{Barbulescu} {et~al.}(2019){Barbulescu}, {Ruderman}, {Van
  Doorsselaere}, \& {Erd{\'e}lyi}}]{barbulescu2019}
{Barbulescu}, M., {Ruderman}, M.~S., {Van Doorsselaere}, T., \& {Erd{\'e}lyi},
  R. 2019, \apj, 870, 108, \dodoi{10.3847/1538-4357/aaf506}

\bibitem[{{Berger} {et~al.}(2008){Berger}, {Shine}, {Slater}, {Tarbell},
  {Title}, {Okamoto}, {Ichimoto}, {Katsukawa}, {Suematsu}, {Tsuneta}, {Lites},
  \& {Shimizu}}]{berger2008}
{Berger}, T.~E., {Shine}, R.~A., {Slater}, G.~L., {et~al.} 2008, \apjl, 676,
  L89, \dodoi{10.1086/587171}

\bibitem[{{Berger} {et~al.}(2010){Berger}, {Slater}, {Hurlburt}, {Shine},
  {Tarbell}, {Title}, {Lites}, {Okamoto}, {Ichimoto}, {Katsukawa}, {Magara},
  {Suematsu}, \& {Shimizu}}]{berger2010}
{Berger}, T.~E., {Slater}, G., {Hurlburt}, N., {et~al.} 2010, \apj, 716, 1288,
  \dodoi{10.1088/0004-637X/716/2/1288}

\bibitem[{Chandrasekhar(1961)}]{chandrasekhar1961}
Chandrasekhar, S. 1961, Hydrodynamic and Hydromagnetic Stability (Clarendon
  Press)

\bibitem[{{Chen} {et~al.}(2020){Chen}, {Zhang}, {De Pontieu}, {Ma}, {Kliem}, \&
  {Priest}}]{chen2020}
{Chen}, H., {Zhang}, J., {De Pontieu}, B., {et~al.} 2020, \apj, 899, 19,
  \dodoi{10.3847/1538-4357/ab9cad}

\bibitem[{Cheung {et~al.}(2015)Cheung, Boerner, Schrijver, Testa, Chen, Peter,
  \& Malanushenko}]{cheung2015}
Cheung, M., Boerner, P., Schrijver, C., {et~al.} 2015, The Astrophysical
  Journal, 807, \dodoi{10.1088/0004-637X/807/2/143}

\bibitem[{{Colgan} {et~al.}(2008){Colgan}, {Abdallah}, {Sherrill}, {Foster},
  {Fontes}, \& {Feldman}}]{colgan2008}
{Colgan}, J., {Abdallah}, J., J., {Sherrill}, M.~E., {et~al.} 2008, \apj, 689,
  585, \dodoi{10.1086/592561}

\bibitem[{{De Pontieu} {et~al.}(2014){De Pontieu}, {Title}, {Lemen}, {Kushner},
  {Akin}, {Allard}, {Berger}, {Boerner}, {Cheung}, {Chou}, {Drake}, {Duncan},
  {Freeland}, {Heyman}, {Hoffman}, {Hurlburt}, {Lindgren}, {Mathur}, {Rehse},
  {Sabolish}, {Seguin}, {Schrijver}, {Tarbell}, {W{\"u}lser}, {Wolfson},
  {Yanari}, {Mudge}, {Nguyen-Phuc}, {Timmons}, {van Bezooijen}, {Weingrod},
  {Brookner}, {Butcher}, {Dougherty}, {Eder}, {Knagenhjelm}, {Larsen},
  {Mansir}, {Phan}, {Boyle}, {Cheimets}, {DeLuca}, {Golub}, {Gates}, {Hertz},
  {McKillop}, {Park}, {Perry}, {Podgorski}, {Reeves}, {Saar}, {Testa}, {Tian},
  {Weber}, {Dunn}, {Eccles}, {Jaeggli}, {Kankelborg}, {Mashburn}, {Pust},
  {Springer}, {Carvalho}, {Kleint}, {Marmie}, {Mazmanian}, {Pereira}, {Sawyer},
  {Strong}, {Worden}, {Carlsson}, {Hansteen}, {Leenaarts}, {Wiesmann},
  {Aloise}, {Chu}, {Bush}, {Scherrer}, {Brekke}, {Martinez-Sykora}, {Lites},
  {McIntosh}, {Uitenbroek}, {Okamoto}, {Gummin}, {Auker}, {Jerram}, {Pool}, \&
  {Waltham}}]{depontieu2014}
{De Pontieu}, B., {Title}, A.~M., {Lemen}, J.~R., {et~al.} 2014, \solphys, 289,
  2733, \dodoi{10.1007/s11207-014-0485-y}

\bibitem[{{D{\'\i}az-Su{\'a}rez} \& {Soler}(2021)}]{diaz2021}
{D{\'\i}az-Su{\'a}rez}, S., \& {Soler}, R. 2021, \aap, 648, A22,
  \dodoi{10.1051/0004-6361/202040161}

\bibitem[{{Hasegawa}(1975)}]{hasegawa1975}
{Hasegawa}, A. 1975, Springer Verlag Springer Series on Physics Chemistry
  Space, 8

\bibitem[{{Hasegawa} {et~al.}(2009){Hasegawa}, {Retin{\`o}}, {Vaivads},
  {Khotyaintsev}, {Andr{\'e}}, {Nakamura}, {Teh}, {Sonnerup}, {Schwartz},
  {Seki}, {Fujimoto}, {Saito}, {R{\`e}me}, \& {Canu}}]{hasegawa2009}
{Hasegawa}, H., {Retin{\`o}}, A., {Vaivads}, A., {et~al.} 2009, Journal of
  Geophysical Research (Space Physics), 114, A12207,
  \dodoi{10.1029/2009JA014042}

\bibitem[{{Heinzel} \& {Shibata}(2018)}]{heinzel2018}
{Heinzel}, P., \& {Shibata}, K. 2018, \apj, 859, 143,
  \dodoi{10.3847/1538-4357/aabe78}

\bibitem[{{Hillier} {et~al.}(2019){Hillier}, {Barker}, {Arregui}, \&
  {Latter}}]{hillier2019}
{Hillier}, A., {Barker}, A., {Arregui}, I., \& {Latter}, H. 2019, \mnras, 482,
  1143, \dodoi{10.1093/mnras/sty2742}

\bibitem[{{Hillier} {et~al.}(2012){Hillier}, {Berger}, {Isobe}, \&
  {Shibata}}]{hillier2012}
{Hillier}, A., {Berger}, T., {Isobe}, H., \& {Shibata}, K. 2012, \apj, 746,
  120, \dodoi{10.1088/0004-637X/746/2/120}

\bibitem[{{Hillier} {et~al.}(2011){Hillier}, {Isobe}, {Shibata}, \&
  {Berger}}]{hillier2011}
{Hillier}, A., {Isobe}, H., {Shibata}, K., \& {Berger}, T. 2011, \apjl, 736,
  L1, \dodoi{10.1088/2041-8205/736/1/L1}

\bibitem[{Howson {et~al.}(2021)Howson, De~Moortel, \& Pontin}]{howson2021}
Howson, T., De~Moortel, I., \& Pontin, D. 2021, A\&A, 656, A112,
  \dodoi{10.1051/0004-6361/202141620}

\bibitem[{{Imada} \& {Zweibel}(2012)}]{imada2012}
{Imada}, S., \& {Zweibel}, E.~G. 2012, \apj, 755, 93,
  \dodoi{10.1088/0004-637X/755/2/93}

\bibitem[{Ishikawa {et~al.}(2017)Ishikawa, Glesener, Krucker, Christe,
  Buitrago-Casas, Narukage, \& Vievering}]{ishikawa2017}
Ishikawa, S.-N., Glesener, L., Krucker, S., {et~al.} 2017, Nature Astronomy, 1,
  771–774, \dodoi{10.1038/s41550-017-0269-z}

\bibitem[{Klimchuk(2006)}]{Klimchuk2006}
Klimchuk, J.~A. 2006, Solar Physics, 234, 41–77,
  \dodoi{10.1007/s11207-006-0055-z}

\bibitem[{{Kriginsky} {et~al.}(2021){Kriginsky}, {Oliver}, {Antolin},
  {Kuridze}, \& {Freij}}]{kriginsky2021}
{Kriginsky}, M., {Oliver}, R., {Antolin}, P., {Kuridze}, D., \& {Freij}, N.
  2021, \aap, 650, A71, \dodoi{10.1051/0004-6361/202140611}

\bibitem[{Kuridze {et~al.}(2016)Kuridze, Zaqarashvili, Henriques, Mathioudakis,
  Keenan, \& Hanslmeier}]{kuridze2016}
Kuridze, D., Zaqarashvili, T.~V., Henriques, V., {et~al.} 2016, The
  Astrophysical Journal, 830, 133, \dodoi{10.3847/0004-637x/830/2/133}

\bibitem[{{Landi} \& {Reale}(2013)}]{landi2013}
{Landi}, E., \& {Reale}, F. 2013, \apj, 772, 71,
  \dodoi{10.1088/0004-637X/772/1/71}

\bibitem[{{Lemen} {et~al.}(2012){Lemen}, {Title}, {Akin}, {Boerner}, {Chou},
  {Drake}, {Duncan}, {Edwards}, {Friedlaender}, {Heyman}, {Hurlburt}, {Katz},
  {Kushner}, {Levay}, {Lindgren}, {Mathur}, {McFeaters}, {Mitchell}, {Rehse},
  {Schrijver}, {Springer}, {Stern}, {Tarbell}, {Wuelser}, {Wolfson}, {Yanari},
  {Bookbinder}, {Cheimets}, {Caldwell}, {Deluca}, {Gates}, {Golub}, {Park},
  {Podgorski}, {Bush}, {Scherrer}, {Gummin}, {Smith}, {Auker}, {Jerram},
  {Pool}, {Soufli}, {Windt}, {Beardsley}, {Clapp}, {Lang}, \&
  {Waltham}}]{lemen2012}
{Lemen}, J.~R., {Title}, A.~M., {Akin}, D.~J., {et~al.} 2012, \solphys, 275,
  17, \dodoi{10.1007/s11207-011-9776-8}

\bibitem[{Li {et~al.}(2018)Li, Zhang, Yang, Hou, \& Erdélyi}]{li2018}
Li, X., Zhang, J., Yang, S., Hou, Y., \& Erdélyi, R. 2018, Scientific Reports,
  8, \dodoi{10.1038/s41598-018-26581-4}

\bibitem[{{Moore} {et~al.}(2016){Moore}, {Nykyri}, \& {Dimmock}}]{moore2016}
{Moore}, T.~W., {Nykyri}, K., \& {Dimmock}, A.~P. 2016, Nature Physics, 12,
  1164, \dodoi{10.1038/nphys3869}

\bibitem[{{Moschou} {et~al.}(2015){Moschou}, {Keppens}, {Xia}, \&
  {Fang}}]{moschou2015}
{Moschou}, S.~P., {Keppens}, R., {Xia}, C., \& {Fang}, X. 2015, Advances in
  Space Research, 56, 2738, \dodoi{10.1016/j.asr.2015.05.008}

\bibitem[{{Nakariakov} \& {Ofman}(2001)}]{nakariakov2001}
{Nakariakov}, V.~M., \& {Ofman}, L. 2001, \aap, 372, L53,
  \dodoi{10.1051/0004-6361:20010607}

\bibitem[{{Nykyri} \& {Otto}(2001)}]{nykyri2001}
{Nykyri}, K., \& {Otto}, A. 2001, \grl, 28, 3565, \dodoi{10.1029/2001GL013239}

\bibitem[{{Nykyri} {et~al.}(2006){Nykyri}, {Otto}, {Lavraud}, {Mouikis},
  {Kistler}, {Balogh}, \& {R{\`e}me}}]{nykyri2006}
{Nykyri}, K., {Otto}, A., {Lavraud}, B., {et~al.} 2006, Annales Geophysicae,
  24, 2619, \dodoi{10.5194/angeo-24-2619-2006}

\bibitem[{Okamoto {et~al.}(2015)Okamoto, Antolin, Pontieu, Uitenbroek,
  Doorsselaere, \& Yokoyama}]{okamoto2015}
Okamoto, T.~J., Antolin, P., Pontieu, B.~D., {et~al.} 2015, The Astrophysical
  Journal, 809, 71, \dodoi{10.1088/0004-637x/809/1/71}

\bibitem[{{Pagano} {et~al.}(2021){Pagano}, {Antolin}, \&
  {Petralia}}]{pagano2021}
{Pagano}, P., {Antolin}, P., \& {Petralia}, A. 2021, \aap, 656, A141,
  \dodoi{10.1051/0004-6361/202141030}

\bibitem[{{Parker}(1988)}]{parker1988}
{Parker}, E.~N. 1988, \apj, 330, 474, \dodoi{10.1086/166485}

\bibitem[{Priest(2014)}]{priest2014}
Priest, E. 2014, Magnetohydrodynamics of the Sun (Cambridge University Press),
  \dodoi{10.1017/CBO9781139020732}

\bibitem[{{Reale} {et~al.}(2011){Reale}, {Guarrasi}, {Testa}, {DeLuca},
  {Peres}, \& {Golub}}]{reale2011}
{Reale}, F., {Guarrasi}, M., {Testa}, P., {et~al.} 2011, \apjl, 736, L16,
  \dodoi{10.1088/2041-8205/736/1/L16}

\bibitem[{{Scullion} {et~al.}(2016){Scullion}, {Rouppe van der Voort},
  {Antolin}, {Wedemeyer}, {Vissers}, {Kontar}, \& {Gallagher}}]{scullion2016}
{Scullion}, E., {Rouppe van der Voort}, L., {Antolin}, P., {et~al.} 2016, \apj,
  833, 184, \dodoi{10.3847/1538-4357/833/2/184}

\bibitem[{{Shi} {et~al.}(2021){Shi}, {Van Doorsselaere}, {Guo}, {Karampelas},
  {Li}, \& {Antolin}}]{shi2021}
{Shi}, M., {Van Doorsselaere}, T., {Guo}, M., {et~al.} 2021, \apj, 908, 233,
  \dodoi{10.3847/1538-4357/abda54}

\bibitem[{{Soler} {et~al.}(2019){Soler}, {Terradas}, {Oliver}, \&
  {Ballester}}]{soler2019}
{Soler}, R., {Terradas}, J., {Oliver}, R., \& {Ballester}, J.~L. 2019, \apj,
  871, 3, \dodoi{10.3847/1538-4357/aaf64c}

\bibitem[{{Testa} \& {Reale}(2012)}]{testa2012}
{Testa}, P., \& {Reale}, F. 2012, \apjl, 750, L10,
  \dodoi{10.1088/2041-8205/750/1/L10}

\bibitem[{{Testa} {et~al.}(2013){Testa}, {De Pontieu}, {Mart{\'\i}nez-Sykora},
  {DeLuca}, {Hansteen}, {Cirtain}, {Winebarger}, {Golub}, {Kobayashi},
  {Korreck}, {Kuzin}, {Walsh}, {DeForest}, {Title}, \& {Weber}}]{testa2013}
{Testa}, P., {De Pontieu}, B., {Mart{\'\i}nez-Sykora}, J., {et~al.} 2013,
  \apjl, 770, L1, \dodoi{10.1088/2041-8205/770/1/L1}

\bibitem[{{Testa} {et~al.}(2014){Testa}, {De Pontieu}, {Allred}, {Carlsson},
  {Reale}, {Daw}, {Hansteen}, {Martinez-Sykora}, {Liu}, {DeLuca}, {Golub},
  {McKillop}, {Reeves}, {Saar}, {Tian}, {Lemen}, {Title}, {Boerner},
  {Hurlburt}, {Tarbell}, {Wuelser}, {Kleint}, {Kankelborg}, \&
  {Jaeggli}}]{testa2014}
{Testa}, P., {De Pontieu}, B., {Allred}, J., {et~al.} 2014, Science, 346,
  1255724, \dodoi{10.1126/science.1255724}

\bibitem[{{Tomczyk} {et~al.}(2007){Tomczyk}, {McIntosh}, {Keil}, {Judge},
  {Schad}, {Seeley}, \& {Edmondson}}]{tomczyk2007}
{Tomczyk}, S., {McIntosh}, S.~W., {Keil}, S.~L., {et~al.} 2007, Science, 317,
  1192, \dodoi{10.1126/science.1143304}

\bibitem[{Uchida \& Kaburaki(1974)}]{uchida1974}
Uchida, Y., \& Kaburaki, O. 1974, Solar Physics, 35, 451–466,
  \dodoi{10.1007/bf00151968}

\bibitem[{{Uzdensky}(2007)}]{uzdensky2007}
{Uzdensky}, D.~A. 2007, \apj, 671, 2139, \dodoi{10.1086/522915}

\bibitem[{{Van Doorsselaere} {et~al.}(2009){Van Doorsselaere}, {Birtill}, \&
  {Evans}}]{vandoorsselaere2009}
{Van Doorsselaere}, T., {Birtill}, D.~C.~C., \& {Evans}, G.~R. 2009, \aap, 508,
  1485, \dodoi{10.1051/0004-6361/200912753}

\bibitem[{{Van Doorsselaere} {et~al.}(2020){Van Doorsselaere}, {Srivastava},
  {Antolin}, {Magyar}, {Vasheghani Farahani}, {Tian}, {Kolotkov}, {Ofman},
  {Guo}, {Arregui}, {De Moortel}, \& {Pascoe}}]{vandoorsselaere2020}
{Van Doorsselaere}, T., {Srivastava}, A.~K., {Antolin}, P., {et~al.} 2020,
  \ssr, 216, 140, \dodoi{10.1007/s11214-020-00770-y}

\bibitem[{{Verwichte} {et~al.}(2009){Verwichte}, {Aschwanden}, {Van
  Doorsselaere}, {Foullon}, \& {Nakariakov}}]{verwichte2009}
{Verwichte}, E., {Aschwanden}, M.~J., {Van Doorsselaere}, T., {Foullon}, C., \&
  {Nakariakov}, V.~M. 2009, \apj, 698, 397, \dodoi{10.1088/0004-637X/698/1/397}

\bibitem[{{Verwichte} {et~al.}(2004){Verwichte}, {Nakariakov}, {Ofman}, \&
  {Deluca}}]{verwichte2004}
{Verwichte}, E., {Nakariakov}, V.~M., {Ofman}, L., \& {Deluca}, E.~E. 2004,
  \solphys, 223, 77, \dodoi{10.1007/s11207-004-0807-6}

\bibitem[{{Wentzel}(1979)}]{wentzel1979}
{Wentzel}, D.~G. 1979, \apj, 233, 756, \dodoi{10.1086/157437}

\bibitem[{{Wyper} {et~al.}(2018){Wyper}, {DeVore}, \& {Antiochos}}]{wyper2018}
{Wyper}, P.~F., {DeVore}, C.~R., \& {Antiochos}, S.~K. 2018, \apj, 852, 98,
  \dodoi{10.3847/1538-4357/aa9ffc}

\bibitem[{{Xia} {et~al.}(2017){Xia}, {Keppens}, \& {Fang}}]{xia2017}
{Xia}, C., {Keppens}, R., \& {Fang}, X. 2017, \aap, 603, A42,
  \dodoi{10.1051/0004-6361/201730660}

\bibitem[{{Yuan} {et~al.}(2019){Yuan}, {Shen}, {Liu}, {Li}, {Feng}, \&
  {Keppens}}]{yuan2019}
{Yuan}, D., {Shen}, Y., {Liu}, Y., {et~al.} 2019, \apjl, 884, L51,
  \dodoi{10.3847/2041-8213/ab4bcd}

\bibitem[{{Zaqarashvili} {et~al.}(2015){Zaqarashvili}, {Zhelyazkov}, \&
  {Ofman}}]{zaqarashvili2015}
{Zaqarashvili}, T.~V., {Zhelyazkov}, I., \& {Ofman}, L. 2015, \apj, 813, 123,
  \dodoi{10.1088/0004-637X/813/2/123}

\bibitem[{{Zhelyazkov} {et~al.}(2015){Zhelyazkov}, {Zaqarashvili}, \&
  {Chandra}}]{zhelyazkov2015}
{Zhelyazkov}, I., {Zaqarashvili}, T.~V., \& {Chandra}, R. 2015, \aap, 574, A55,
  \dodoi{10.1051/0004-6361/201424793}

\end{thebibliography}
\end{document}